\documentclass{svmult}
\usepackage{makeidx}         
\usepackage{graphicx}        
\usepackage{multicol}        
\usepackage[bottom]{footmisc}

\makeindex             

\usepackage{epsfig}
\usepackage{graphics}
\usepackage{dcolumn}
\usepackage{bm}

\def\cm-1{cm$^{-1}$}
\def\scco{Sr$_{14-x}$Ca$_{x}$Cu$_{24}$O$_{41}$\,}
\def\slcco{(Sr,La)$_{14-x}$Ca$_{x}$Cu$_{24}$O$_{41}$\,}
\def\scbo{SrCu$_{2}$(BO$_{3}$)$_{2}$\,}
\def\navo{NaV$_{2}$O$_{5}$\,}
\def\naxvo{Na$_{x}$V$_{2}$O$_{5}$\,}
\def\nacavo{Na$_{1-x}$Ca$_{x}$V$_{2}$O$_{5}$\,}

\begin{document}

\title*{Electronic Properties of $\alpha'$-\navo}
\author{A. Gozar\inst{1,2} \and G. Blumberg\inst{1,*}}
\institute{
Bell Laboratories, Lucent Technologies, Murray Hill, NJ 07974, USA \and 
University of Illinois at Urbana-Champaign, Urbana, IL 61801, USA
\texttt{
\begin{center}
    (\textit{Frontiers in Magnetic Materials} (Ed. A.V. Narlikar), 
    Spinger-Verlag 2005, pp. 697-734).
\end{center}
}} 
%
%
\maketitle

\section{General properties of $\alpha'-$\navo and motivation for a spectroscopic study}

{\bf Introduction -- }
$\alpha'-$\navo is one of the several phases in the class of Na$_{x}$V$_{2}$O$_{5}$ systems \cite{IsobeJPSJ96} and until now it is by far the most studied of them.
Since 1996 this compound (denoted in the following simply by \navo) has received considerable attention because it was thought to be the second realization, after CuGeO$_{3}$, of a quasi-one dimensional (1D) \emph{inorganic} material displaying a spin-Peierls (SP) transition.
The interest was justified given the scarcity of inorganic materials having this property, which is quite interesting especially for the scientific community working in the field of low dimensional quantum spin systems.
However, it turned out that the physics of \navo is more complicated and intriguing than that and the degrees of freedom involved are not only the ones describing the spins and the lattice.

What is a SP transition?
We discussed in Chapter \cite{ChapterSCCO} the general properties of a Peierls distortion which is a transition to charge density wave state.
This means that below some temperature T$_{P}$ the crystal gets distorted and the electronic density acquires a periodic spatial modulation, a process during which the loss in elastic energy is compensated by the gain in the kinetic energy of the electrons.
A crucial role is played by the nesting properties of the Fermi surface (i.e. the property that enables one to connect points of the Fermi surface by wavevector characteristic of other excitations, in this case phonons) which makes low dimensional systems especially susceptible to such an instability.
A pure SP transition is one in which the lattice distortion is caused by the magneto-elastic coupling, the gain in energy in this case being related to the spin degrees of freedom \cite{PyttePRB74}.
In other words, it is a lattice instability driven by the magnetic interactions.
This phenomenon leads to the formation of a spin-singlet ($S = 0$) ground state and the opening of a spin-gap in the magnetic excitation spectrum, i.e. a finite energy is required to excite the system from its ground state to lowest triplet ($S = 1$) state. 
A signature of a SP state is thus an isotropic activated temperature dependence in the uniform magnetic susceptibility below the transition at T$_{SP}$.
In addition, as opposed to an usual Peierls transition, the direct participation of the spins leads to specific predictions for the dependence of T$_{SP}$ on external magnetic field $H$ \cite{BulaevskiiSSC78}.
This has to do with the fact that in the spin case the filling factor of the electronic band and accordingly the magnitude of the nesting wavevectors can be varied continuously by a magnetic field.
This statement should not be taken \emph{ad litteram}, but in the sense that the magnetic problem can be mapped onto a fermion like system by using a transformation of the spin operators, the magnetic field playing the role of the chemical potential, see Refs.~\cite{PyttePRB74,BulaevskiiSSC78} for more details.

\begin{figure}[t]
\centerline{
\epsfig{figure=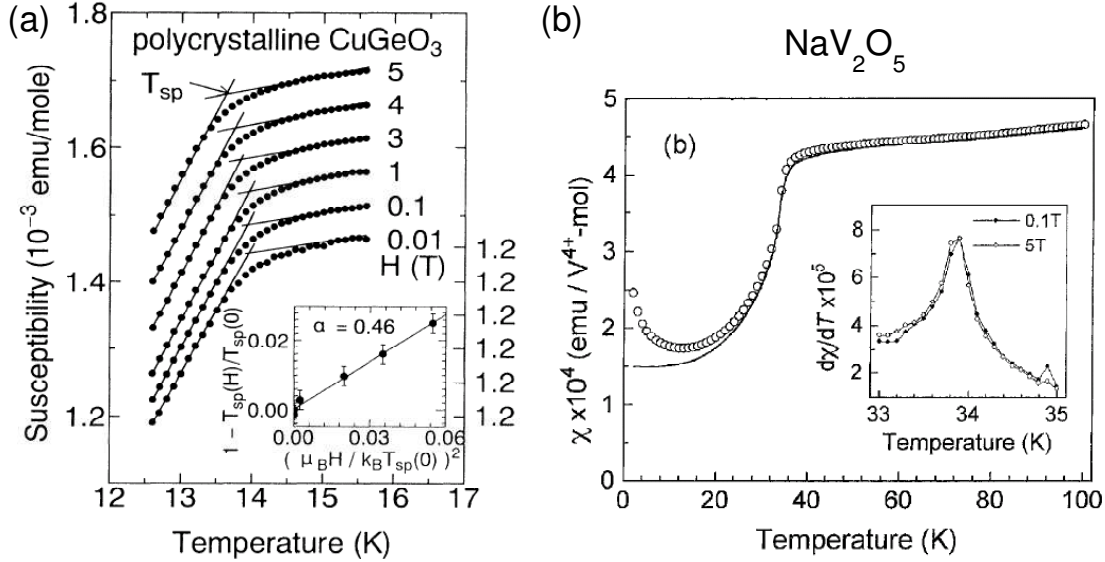,width=5.75in}
}
\caption{
(a) Magnetic susceptibility of polycrystalline CuGeO$_{3}$ as a function of temperature for several values of external magnetic fields.
Inset: the reduced temperature $1 - T_{SP} (H) / T_{SP} (H = 0)$ as a function of the dimensionless parameter $\mu_{B} H / k_{B} T_{SP} (H = 0)$; the proportionality coefficient is 0.46.
Data from Ref.~\cite{HasePRL93}.
(b) Temperature dependent susceptibility in powder samples of \navo.
The inset shows the derivative $d \chi(T) / d T$ as function of temperature for 0.1 and 5~T.
Data from Ref.~\cite{IsobeJPSJ96}.
}
\label{f51}
\end{figure}

So what is the difference between CuGeO$_{3}$ and \navo?
In the former, the SP nature of the transition observed around 14~K was inferred from the exponential drop in the magnetic susceptibility below this temperature and especially from the dependence of T$_{SP}$ on an applied external field \cite{HasePRL93}, see Fig.~\ref{f51}a.
As predicted by theory, it was found that the field dependence was quadratic, $1 - T_{SP}(H) / T_{SP}(H = 0) = \alpha \cdot [ ( \mu_{B} H ) / (k_{B} T_{SP} (H = 0) ) ]^{2}$.
Moreover, the experimental proportionality factor $\alpha = 0.46$ was also in very good agreement with the theoretical mean field value \cite{BulaevskiiSSC78}.
Finally, neutron scattering experiments and detailed investigations of the field-temperature phase diagram confirmed the previous SP interpretation~\cite{BoucherJdP96}.

As for \navo, many common properties with CuGeO$_{3}$ were observed.
In 1996 an isotropic activated behavior below T$_{c} = 34$~K was observed by Isobe and Ueda \cite{IsobeJPSJ96}, see Fig.~\ref{f51}b.
Soon after that there appeared the first report of X-ray and inelastic neutron scattering (INS) data in \navo \cite{FujiiJPSJ97}. 
The authors reported the existence below T$_{c} \approx 35$~K of superlattice peaks due to a modulation given by $k = (2 \pi / 2a, 2 \pi / 2 b, 2 \pi / 4c)$.
This wavevector characterizes distortions corresponding to unit cell doubling along the $a$ and $b$ axes and quadrupling along the $c$ direction.
In the same paper, INS results showed that the transition is accompanied by the opening of a spin gap whose value was estimated to be $\Delta_{S} \approx 9.8$~meV.
So far these results are qualitatively similar to the ones in CuGeO$_{3}$ but it is worth noting that a value $2 \Delta_{S} / k_{B} T_{c}$ of approximately 6.5 already suggests a departure from the mean field value 3.52.
The interpretation in terms of a simple SP transition became more questionable when experiments in magnetic fields were performed.
Note the inset of Fig.~\ref{f51} where the susceptibility derivative is shown for two values of the external field.  
Specific heat measurements in chemically well characterized crystalline samples of Na$_{x}$V$_{2}$O$_{5}$ with $0.95 \leq x \leq 1$ in fields up to 16~T revealed a single $\lambda$-shaped anomaly at T$_{c} = 33.5$~K and an isotropic decrease of the anomaly with $\Delta T_{c} \propto H^{2}$ \cite{SchnellePRB99}.
However, the proportionality factor, the parameter $\alpha$ introduced in the previous paragraph, was found to be only 20\% of the mean field value \cite{BulaevskiiSSC78}.
Similar measurements aided by magnetic susceptibility data showed a $T_{SP} (H = 0) - T_{SP} (H = 14 T) \approx 0.15$~K, about a factor of 7 smaller than the expected value \cite{DamascelliThesis99}.
Heat capacity measurements also pointed out incompatibilities with a simple SP transition \cite{HembergerEL98}.
The authors of this paper could not reconcile the mean field predictions of the magnetic contribution to the heat capacity with the experimental data: the specific heat jump $\delta C / k_{B} T_{c}$ was found to be a factor of 20 higher than 1.43, which is the mean-field value, if the linear contribution $C = \gamma T$ at high temperatures was fixed to the theoretical expectations for a 1D $S = 1/2$ antiferromagnetic (AF) chain.
\begin{figure}[t]
\centerline{
\epsfig{figure=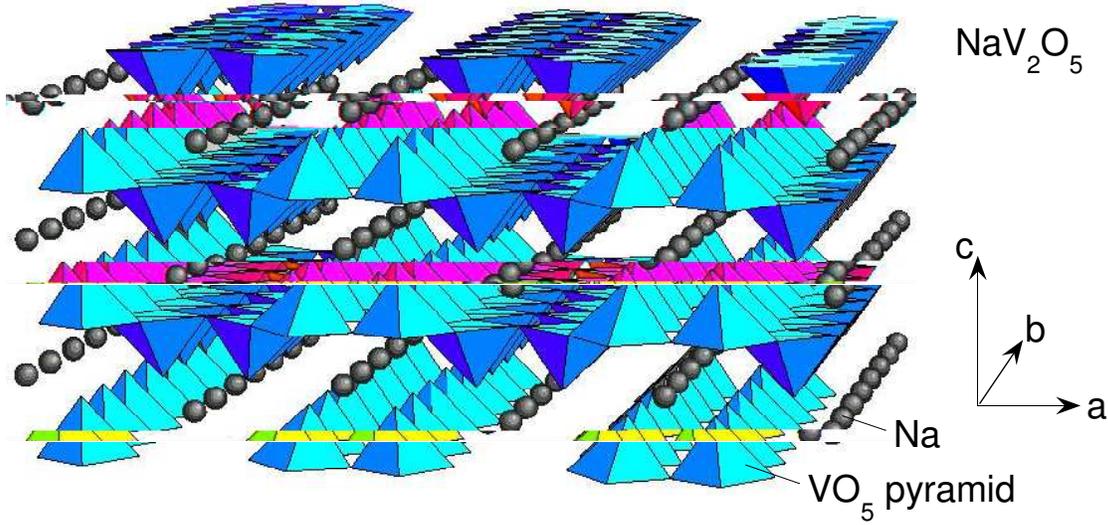,width=5.75in}
}
\caption{
The 3D crystal structure of \navo (adapted from Ref.~\cite{DamascelliThesis99}).
Na atoms form 1D chains along the $b$-axis and pairs of corner sharing VO$_{5}$ pyramids pointing in the same direction form V$_{2}$O$_{5}$ ladder rungs.
The legs of these ladders run along the $b$-axis, see also Fig.~\ref{f53}.
}
\label{f52}
\end{figure}

The discussion above gives us a feeling about some of the important parameters and energy scales one has in mind when discussing the properties of \navo.
The connection made with the 1D AF $S = 1/2$ chain will get support from the description of the crystal structure which is done in the following and will become more clear also in the next section where other properties of \navo in the high temperature phase are discussed.
However, we conclude from what has been said that in order to explain quantitatively the experimental findings one has to take into account other degrees of freedom, contrary to the initial belief that at T$_{c}$ one deals solely with a magnetically driven phase transition.
It has become clear that the coupling of the lattice to the charge degrees of freedom is crucial for understanding the details of what happens at T$_{c}$.
The slow advance in this field was also due to the difficulty in the interpretation of X-ray diffraction data and because of a wrong identification of the crystal structure in an early study \cite{CarpyACB75} which generated a plethora of subsequent papers which were at odds with each other.
Diffraction anomalous fine structure spectroscopy (also called resonant X-ray diffraction, an example of which was discussed in Chapter \cite{ChapterSCCO} in connection to the existence of a density wave order in doped two-leg ladder structures) finally brought more understanding of the nature of the charge ordering in \navo, see Ref.~\cite{GrenierPRB02JolyPRB03} and citations therein.
However, in spite of years of intensive investigation of both the high and low temperature phases of \navo, no microscopic model has successfully explained all the features of the transition and, as we will discuss in more detail, the nature of prominent excitations seen in spectroscopic experiments as well as the light coupling mechanisms to the collective spin and/or charge excitations is still far from being clear.

\begin{figure}[t]
\centerline{
\epsfig{figure=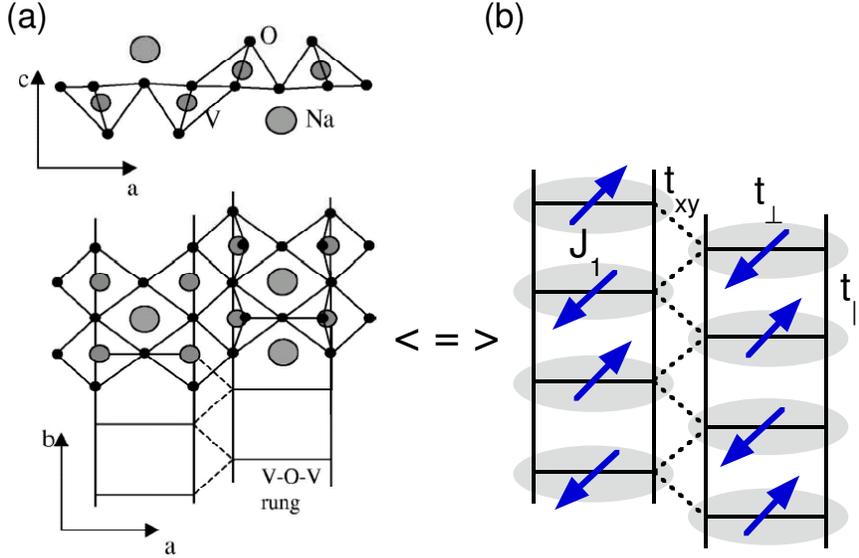,width=4.5in}
}
\caption{
(a) Schematic representation of the \navo in the $(010)$ (upper figure) and $(001)$ (lower figure) planes (from Ref.~\cite{KonstantinovicPRB02}).
(b) Cartoon with the spin structure in the high temperature phase.
The structure can be mapped on a weakly interacting array of quasi 1D $S = 1/2$ spin chains. 
$J_{1}$ is the superexchange between neighboring spins along the $b$-axis and $t_{\perp}$, $t_{\parallel}$ and $t_{xy}$ are the rung, leg and inter-ladder hopping parameters.
}
\label{f53}
\end{figure}

{\bf Structure and electronic properties -- }
The crystal structure of \navo is shown in Fig.~\ref{f52}.
It contains Vanadium-Oxygen planes stacked along the $c$-axis and separated by Na chains.
Each V-O plane is formed from pairs of edge sharing VO$_{5}$ pyramids running along the $b$-axis (the vertices of the pyramids in every such pair of chains are pointing in opposite direction along the $c$-axis), each pair being connected in turn to a neighboring one by corner sharing pyramids.
In the $(ab)$ plane, see also Fig.~\ref{f53}, one can identify two-leg ladder (2LL) units, very similar to the ones found in \scco compounds.
One can look for instance at a pair of adjacent rows of VO$_{5}$ pyramids pointing in the same direction along the $c$-axis.
The rungs of the ladder, oriented along the $a$-axis are formed by a pair two V ions bridged by an O atom which is at the common corner of two pyramids.

Can this structure be mapped onto an array of quasi-1D electronic units?
A valence counting for Na$^{+}$V$_{2}^{4.5+}$O$_{5}^{2-}$ shows that the formal V valence is 4.5+.
The electronic configuration of a neutral V atom is $3d^{3}4s^{2}$, so a V$^{5+}$ ion has a closed shell configuration while the V$^{4+}$ has one $3d$ electron on the upper shell.
On the average there is one electron shared by two V ions and this makes the 2LL's to be at quarter filling factor. 
The initially proposed non-centrosymmetric $P2_{1}mn$ crystal symmetry group at 300~K \cite{CarpyACB75} allowed for two inequivalent V positions, interpreted as magnetic V$^{4+}$ and non-magnetic V$^{5+}$ sites.
The magnetic properties were thought to be determined by rows of $S = 1/2$ V$^{4+}$ ions along a leg of a ladder (with the other one remaining magnetically inert), each row of V$^{4+}$ ions being only weakly coupled to to the adjacent one due to large separation between them. 
It is now accepted based on X-ray diffraction studies that above T$_{c}$ the correct space group is the centrosymmetric $Pmmn$ group \cite{SmolinskiPRL98,MeetsmaACC98,LudeckePRL99}.
This is important in what regards the number of distinct of V atoms existent in the high temperature phase: the inversion center and the the two vertical mirror planes (i. e. perpendicular to the $(ab)$ plane) imply that there is only one type of V atoms with an effective valence of 4.5+.
The fact that \navo is in a mixed-valence state is further supported by nuclear magnetic resonance (NMR) studies which confirmed the average oxidation state of all vanadium atoms to be V$^{4.5+}$ \cite{OhamaPRB99OhamaJPSJ00}.  

Accordingly, one can think about the ladder structure of \navo as having on each rung an electron equally shared by pairs of V atoms forming the rungs.
Susceptibility data also indicate that the coupling between the spins of these electrons is antiferromagnetic (AF) \cite{IsobeJPSJ96}.  
The local symmetry splits the five-fold degeneracy of the $d$ shell and makes the $d_{xy}$ band (the zero's of the wave function are along the $a$ and $b$ axes) the lowest in energy compared to the bands generated by the other orbitals \cite{SmolinskiPRL98}.
Consequently, the $d_{xy}$ orbital is the relevant atomic state for the analysis of the low energy charge excitations.
How about the quasi one dimensionality?
A density functional calculation \cite{SmolinskiPRL98} of the energy bands and their mapping on tight-binding models yields for the hopping terms the following values: a rung hopping $t_{\perp} \approx 0.38$~eV, a leg hopping $t_{\parallel} \approx 0.17$~eV and a inter-ladder hopping $t_{xy} \approx 0.012$~eV.
A difference of more than an order of magnitude between the intra-ladder and inter-ladder wavefunction overlap is theoretical proof for the quasi-1D nature of the system.
As we will show later in relation to the low temperature phase, the dispersions of the low energy collective excitations along the $a$-axis, which were found to be much smaller than the ones along the leg direction, confirm experimentally this hypothesis.
This is also true at 300~K, as can be inferred from the band dispersions derived from an angle resolved photoemission (ARPES) study \cite{KobayashiPRL98}. 

In a regular Fermi liquid system, the V valence would make this compound metallic.
However, an insulating character inferred from resistivity measurements was confirmed to exist both above and below T$_{c}$ \cite{HembergerEL98}.
This behavior is due to correlation effects \cite{MostovoySSC00andPRB02} and it is explained below within a simple model.
Consider a rung having one electron in a bonding orbital.
The next excitation would be a transition to the antibonding state situated somewhere about $2 t_{\perp}$ above.
Assuming that the on-site Coulomb energy $U$ is infinite, i.e. no two electrons can be found on the same V $d$ shell, an electron can hop only on an empty site of a neighboring rung.
However, this will cost an energy of the order of $2 t_{\perp}$.
The quarter filled ladder becomes thus equivalent to a half filled Hubbard chain with an ``effective" on-rung repulsion $U_{eff} = 2 t_{\perp}$ \cite{MostovoySSC00andPRB02}, explaining in principle why the vanadium ladders are insulating.

{\bf Experimental -- }
In the following we will discuss properties of the high and low temperature phases of \navo, concentrating on the results of our Raman scattering experiments.
The single crystal we measured had the $a \times b \times c$ dimensions around $2 \times 4 \times 0.5$~mm and was grown as described in \cite{JohnstonPRB00}.
Data were taken in a backscattering geometry using linearly and circularly (the latter only at the lowest temperatures in magnetic fields) light from a Kr$^{+}$ laser.
All the data were corrected for the spectral response of the spectrometer.
In addition, the resonance profile at 300~K took into account the change in the optical properties of the material as the excitation energy $\omega_{in}$ was swept from infra-red (IR) to violet.
The incident photons propagated along the $c$-axis and by $(xy)$ we denote a polarization configuration with the incoming photon polarization ${\bf e}_{in} \parallel \hat{x}$ and outgoing polarization ${\bf e}_{out} \parallel \hat{y}$.

\section{Magnetic Raman continuum in the high temperature phase ($T~\geq~34$~K)}

\subsection{Experimental properties: polarization, resonance and temperature dependence}

\begin{figure}[t]
\centerline{
\epsfig{figure=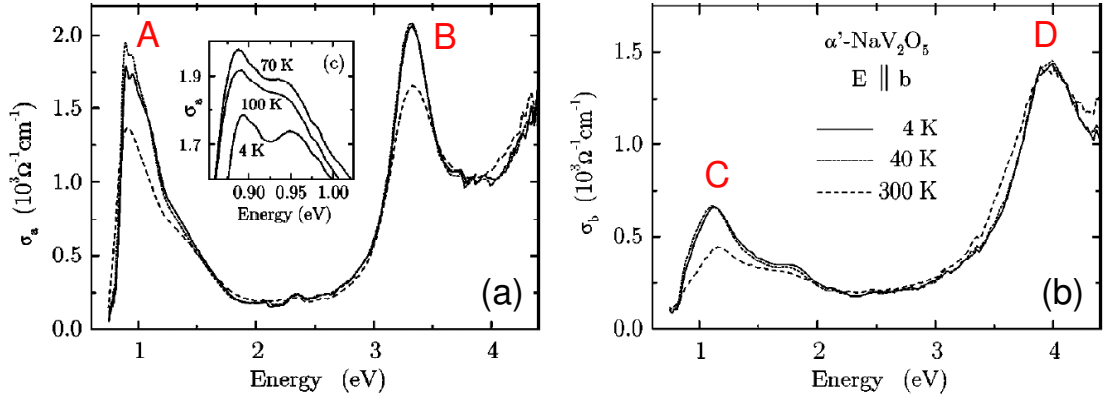,width=5.75in}
}
\caption{
Optical conductivity at 4, 40 and 300~K for electric fields $E \parallel a$ (left) and $E \parallel b$ (right) from Ref.~\cite{PresuraPRB00}.
The letters A, B, C and D mark the main absorption features in the 0~-~4~eV energy range.
}
\label{f54}
\end{figure}

Before illustrating the experimental properties of the observed broad Raman continua, we show the main features of the absorption spectrum in an energy range up to about 4~eV \cite{PresuraPRB00}.
The absorption properties will be discussed in some detail because we will use these results in the analysis of our resonance Raman study.
Optical conductivity data from Ref.~\cite{PresuraPRB00} are shown in Fig.~\ref{f54}.
The spectra show the relevant energy scales involving electron dynamics along and across the legs.
The left panel the electric field is was polarized along the rung direction while in the right panel the polarization was parallel to the legs direction.
The $a$-axis polarized spectra show a strong peak at 0.9~eV (peak A) with a shoulder at 1.4~eV and another peak at 3.3~eV (peak B).
A similar, but blueshifted sequence, is observed for light polarization along the $b$ direction, the energies of the observed excitations being around 1.2 (peak C), 1.9 and 3.9~eV (peak D).
It should be noted that the relevant electronic orbitals to be taken into account when discussing excitations in this energy range are the Vanadium $3d$ and Oxygen $2p$ states. 

In Ref.~\cite{PresuraPRB00} the authors assign peak A to an optical transition from a bonding to an anti-bonding orbital made out of symmetric (and antisymmetric) combinations of $d_{xy}$ orbitals of the two V atoms forming a rung.
This is the fundamental gap of the optical spectrum.
This interpretation was supported on one hand by an ARPES study \cite{KobayashiPRL98} which shows that the top of the bands generated by the O $2p$ orbitals are about 3~eV below the V $3d$ manifold as well as by a band structure calculation \cite{SmolinskiPRL98} and on the other hand, by a study of the evolution of this excitation with Ca doping in \nacavo \cite{PresuraPRB00}.
It has been observed that the spectral weight of this peak decreases linearly with Ca concentration which is consistent with a diminishing intensity due to the fact that with two electrons per rung (Ca has a 2+ valence) the many-body 2-particle state no longer allows a low energy bonding-antibonding transition \cite{DamascelliPRL98}.
The above mentioned Ca dependence was also used to rule out the scenario that the 0.9~eV peak is a result of V $3d \leftrightarrow 3d$ transition, the argument being that such an excitation would be proportional to the number of V$^{4+}$ ions which increases with Ca doping.
If peak A is essentially an on-rung excitation, peak C at 1.1 eV was assigned to an optical excitation involving neighboring rungs \cite{PresuraPRB00}.
The final wave function is different than the one corresponding to peak A because it involves combination of states having one rung with zero electrons and another one being doubly occupied, the red-shift of about 0.3-0.5~eV seen in $a$-axis polarization reflecting thus excitonic effects for the on-rung electron hole pair.

As for peaks B and D, they are thought to arise likely from transitions between O $2p$ and V $3d_{xy}$ manifolds.
Remarkable is also the substantial absorption all the way up to 4~eV and even beyond which explains the black color of the \navo crystals.
We will discuss in the following polarization, resonance of the Raman continuum we observe in the 0~-~200~meV energy range as well as its temperature dependence.
A discussion of our proposed scenario for its origin will be contrasted to other present interpretations.

\begin{figure}[t]
\centerline{
\epsfig{figure=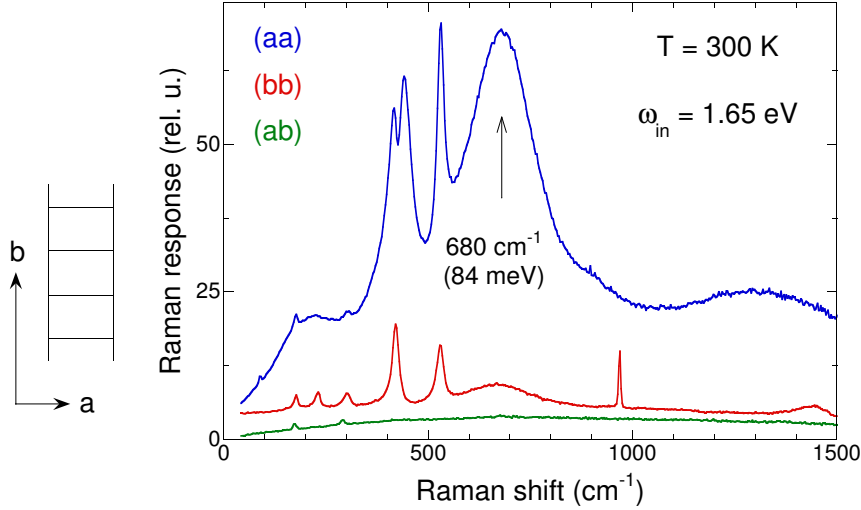,width=4.5in}
}
\caption{
T~=300~K Raman spectra of \navo in $(aa)$, $(bb)$ and $(ab)$ polarizations taken using $\omega_{in} = 1.65$~eV laser excitation energy. 
}
\label{f55}
\end{figure}

{\bf Polarization properties -- }
Fig.~\ref{f55} shows three room temperature Raman spectra taken in $(aa)$, $(bb)$ and $(ab)$ polarizations and using the $\omega_{in} = 1.65$~eV excitation energy.
The $(ab)$ polarized spectrum has the lowest intensity and it is rather featureless.
Next in terms of the overall background intensity comes the spectrum in $(bb)$ polarization.
Several relatively sharp features are seen at 420, 530 and 968~\cm-1 and also some weaker ones at lower energies, 177, 230 and 300~\cm-1.
Besides these modes, there is a broad excitation peaked at 680~\cm-1, but not particularly strong.
The sharp modes are phonons, excitations which were studied intensively both by Raman and IR spectroscopy \cite{GolubchikPopovicPopova97-99}.
The strong continuum which extends all the way from the origin and having a main a peak around 680~\cm-1 becomes the most prominent feature of the Raman spectrum in $(aa)$ polarization.
There are differences in phonons, for instance in this geometry we see two modes close to 420~\cm-1 as opposed to only one in $(bb)$ configuration and also the 968~\cm-1 peak is not present. 
We also note the observation of a smaller shoulder at about 200~\cm-1 and, interestingly, of another excitation around 1320~\cm-1, which is approximately twice the energy of the main peak at 680~\cm-1.

\begin{figure}[t]
\centerline{
\epsfig{figure=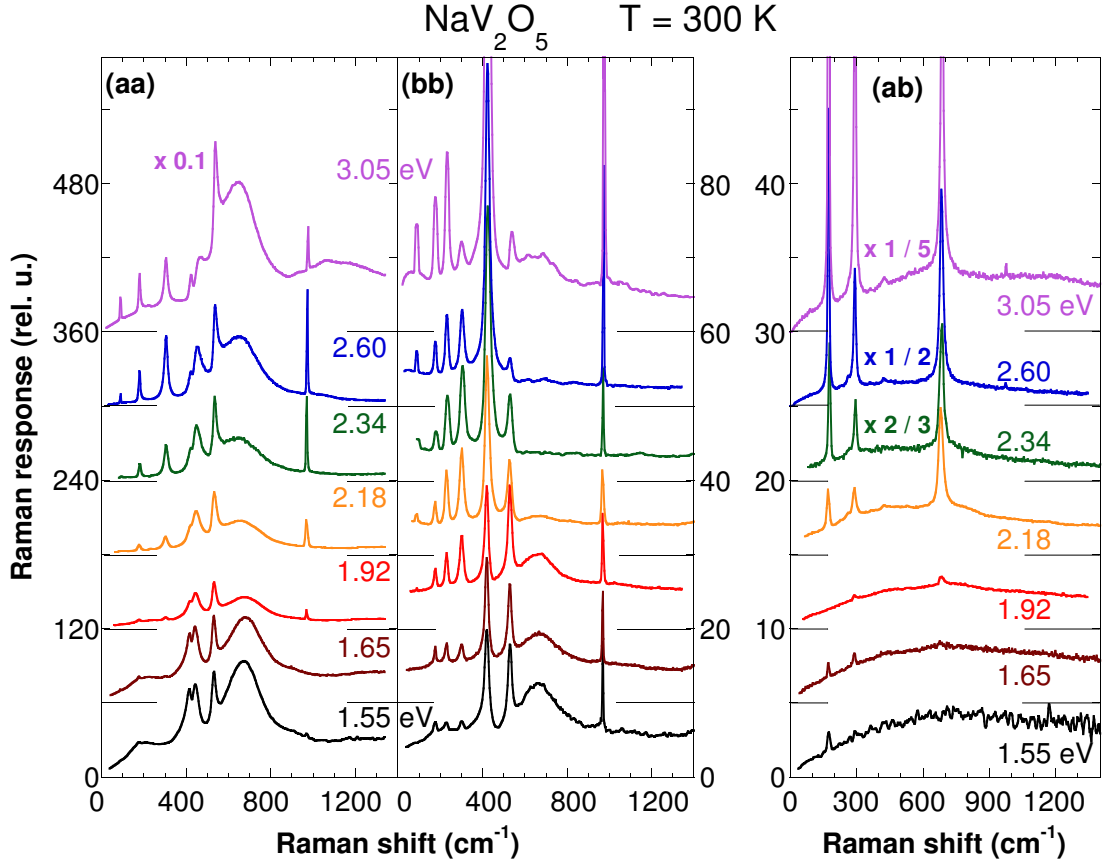,width=5.75in}
}
\caption{
Raman response at $T = 300$~K in $(aa)$ (left panel), $(bb)$ (middle panel) and $(ab)$
(right panel) polarizations for different incoming photon energies.
Note the relative intensity scales for the three polarizations as well as the multiplication factors used for the 3.05~eV spectrum in $(aa)$ polarization and 3.05, 2.60 and 2.34~eV spectra in $(ab)$ configuration.
}
\label{f56}
\end{figure}

{\bf Raman resonant profile at $ T = 300$~K -- }
In Fig.~\ref{f56} we show the dependence of the intensity of this continuum as a function of the energy of the incoming photons $\omega_{in}$.
The resonance behavior is shown for the three polarizations shown in Fig.~\ref{f55}.
Line colors for each of the spectra shown are chosen so that they roughy correspond to the actual color of the laser excitation beam in the visible spectrum. 
Besides the usual spectrometer/detector correction, the spectra shown in this figure were corrected for the optical properties of the crystal.
The absorption coefficients the refraction indexes as well as the transmission at the sample interface were calculated using the real and the imaginary parts of the complex dielectric function provided to us by the authors of Ref.~\cite{PresuraPRB00}.
The spectral weight of the Raman continuum in the $(aa)$ polarized spectra seems to be peaked at the extremities of the excitation range of the Kr$^{+}$ laser, 1.55 and 3.05~eV respectively.
As for the $\omega_{in} = 1.65$~eV spectrum shown in Fig.~\ref{f55}, the broad Raman band is present also in $(bb)$ polarization and the resonant enhancement in this configuration follows roughly the behavior in $(aa)$ geometry.
Obvious signatures of the $(bb)$ polarized continuum can be seen for instance in the $\omega_{in}$~=~1.55, 1.65, 1.92 or 3.05~eV spectra, while for excitation energies corresponding to the yellow, green and blue in the visible spectrum this Raman band is absent.
It is important to note that although in cross polarization we do not observe a similar structure of the continuum as in $(aa)$ and $(bb)$ geometries, i.e. a broad Raman band peaked around 680~\cm-1, the overall intensity of the Raman background is excitation dependent, and a simple inspection of the right panel in Fig.~\ref{f56} suggests that in $(ab)$ configuration we see a slight decrease of the background as $\omega_{in}$ is decrease from 1.55 to 1.92~eV followed by an uprise at the other end of the spectrum, see the multiplication factors used for the 2.34, 2.60 and 3.05~eV.

The large width of the observed continuum makes it a distinct feature compared with the other sharp phononic lines seen in the spectra of Fig.~\ref{f56} and points strongly to its electronic origin.
Since the phonons are not the focus of the present study, we make only a few remarks about these features at room temperature.
The interested reader can consult Raman and IR studies in Refs.~\cite{GolubchikPopovicPopova97-99,KonstantinovicPSS99,KonstantinovicJPCM99} for a detailed analysis of their behavior as a function of polarization, excitation energy and temperature along with some lattice dynamical calculations.
In all three polarizations we observe an enhancement in the Raman intensity of the low energy phonons in the 100~-~400~\cm-1 energy range.
Some of the phonons situated in the region where the electronic Raman band is more intense have a pronounced asymmetric (Fano) shape due to the interaction with the underlying continuum.
The 531~\cm-1 mode (assigned to V-O$_{leg}$ stretching mode \cite{GolubchikPopovicPopova97-99}) is proof for the interaction between the the lattice and the electronic degrees of freedom \cite{KonstantinovicPSS99}, see for instance the spectra corresponding to excitation energies $\omega_{in} \geq 2.34$~eV in $(aa)$ polarization.
\begin{figure}[t]
\centerline{
\epsfig{figure=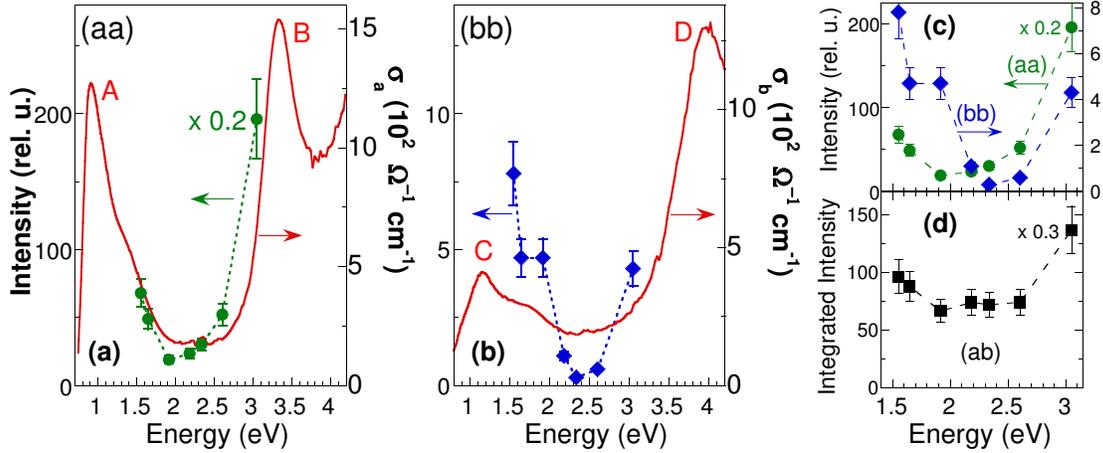,width=5.75in}
}
\caption{
(a) Intensity of the electronic Raman band (extracted from the 680~\cm-1 peak hight) in $(aa)$ polarization as function of the excitation energy $\omega_{in}$ (green solid circles).
Dashed lines are guides for the eye.
The a-axis optical conductivity $\sigma_{a}$ is plotted on the right scale (solid red line).
(b) The b-axis optical conductivity $\sigma_{b}$ is shown by a solid red line, and the resonant profile in $(bb)$ polarization is shown by blue diamonds.
(c) Resonance profiles in $(aa)$ (left scale) and $(bb)$ (right scale) polarizations from panels (a) and (b) but this time plotted together.
The same symbols as in (a) and (b) are used.
Note the blue shift of the profile in $(bb)$ configuration.
(d) The intensity of the continuum in $(ab)$ polarization obtained by integrating the Raman response, $\int \chi''(\omega) \ d\omega$, between 100 and 1200~\cm-1 with the phonons masked.
The optical conductivity data is courtesy of the authors of Ref.~\cite{PresuraPRB00}.
}
\label{f57}
\end{figure}

Fig.~\ref{f57}a-b shows the Raman resonant excitation profile for the 680~\cm-1 band plotted against the optical conductivity data, see Fig.~\ref{f54} and Ref.~\cite{PresuraPRB00}.
The red line is the the optical conductivity $\sigma (\omega)$ and the same notations as in Fig.~\ref{f54} are used for the main four absorption features seen in $a$ and $b$ axes polarized spectra.
The dots represent the intensity of the Raman band extracted from the 680~\cm-1 peak height.
This way of analyzing the data was chosen in the absence of a suitable fitting function for the background and also because of the presence of strong phononic features interacting with the continuum.
The resonant enhancement towards both ends of the visible spectrum remarked in the discussion of Fig.~\ref{f56}, can be clearly observed here.
The efficiency of the inelastic cross section follows very closely the absorption bands, showing that this excitation is always seen 'in resonance'.
Importantly, if the resonance profiles for $(aa)$ and $(bb)$ polarizations are plotted together, see the inset of Fig.~\ref{f57}c, one can observe that the blue shift of the $b$-axis polarized optical peaks with respect to the peaks in the $a$-axis spectra is also reflected in the dependence of the Raman band intensity as a function of excitation frequency.
Although the clean phononic selection rules we see in the data which can be easily checked for instance by comparison of cross and parallel polarized spectra, see Figs.~\ref{f515} and \ref{f516}, are strong reasons to believe that the $(bb)$ continuum is not a result of sample misalignment in conjunction with polarization ``leakage'' from the $(aa)$ spectra, the agreement between the blue shift seen in both the optical spectra and the resonance profile for the two polarizations leaves no doubt that the feature seen in $(bb)$ data is a true Raman signal.

The 680~\cm-1 peak is absent in $(ab)$ polarization, but we still see in this case a signal which is excitation energy dependent.
This signal is not coming from spurious effects related to crystal quality and the complete absence of scattering in the 0-2$\Delta$ energy region in the low temperature phase, where $\Delta = 66$~\cm-1 is the T~=~10~K low temperature spin gap which opens below T$_{c}$~=~34~K, supports this statement. 
In Fig.~\ref{f57}d we show the integrated $(ab)$ polarized Raman response $\chi''(\omega)$ between 100 and 1200~\cm-1 with the sharp phononic features masked.
As was the case for the data in parallel polarizations, we observe again an enhancement in the background for the lowest and highest excitation energies $\omega_{in}$.

According to the interpretation of the optical features seen in Fig.~\ref{f54}, the intermediate electronic states responsible for the resonant behavior involve on-rung bonding-antibonding transitions, inter-rung excitations as well as transitions from the lower O $2p$ states to the V $3d$ bands.
Irrespective of the interpretation, we established so far that the continuum peaked at 680~\cm-1 appears in both $(aa)$ and $(bb)$ polarizations, the peak is not present in cross polarization and is a strongly excitation energy dependent feature, which allows us to clearly see its next overtone for incoming photon frequencies $\omega_{in} \geq 2.6$~eV or $\omega_{in} \leq 1.65$~eV.
\begin{figure}[t]
\centerline{
\epsfig{figure=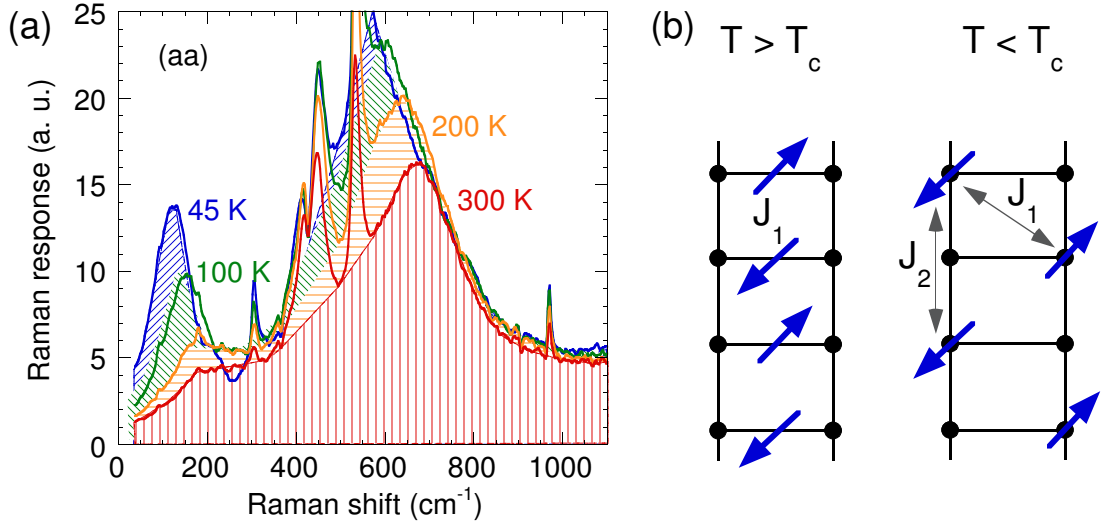,width=5.75in}
}
\caption{
(a) Raman scattering continuum in \navo taken in $(aa)$ polarization at 300, 200, 100 and 45~K using the $\omega_{in} = 1.83$~eV excitation energy.
Note the softening of the band peaked at 680~\cm-1 at room temperature as well as the development of another continuum at frequencies below 200~\cm-1.
The sharp modes are phonons.
(b) Cartoon showing the average distribution of the valence electrons provided by V  atoms (represented by solid black dots) within a given ladder above (left) and below the transition (right).
The picture on the right does not show the additional lattice dimerization and the alternation of the nearest neighbor superexchange $J_{1}$, see the following section for a more detailed discussion of the low temperature phase.
}
\label{f58}
\end{figure}

{\bf Temperature dependence for $34 < T < 300$~K and the role of fluctuations above the critical temperature -- }
We show in Fig.~\ref{f58}a the dependence on temperature of the Raman continuum as we cool down from 300~K to just about the transition.
The data are taken in $(aa)$ polarization and the excitation energy used was $\omega_{in} = 1.83$~eV.
We remark two effects regarding this spectra.
The first is the downshift with cooling of the strong Raman band in the 680~\cm-1 region accompanied by an increase in its spectral weight from the low energy side.
Although the interaction with the 530~\cm-1 phonon may lead to an additional shift compared to its ``bare'' energy, one can observe by simple inspection of the spectra a substantial softening of about 80~\cm-1 with cooling from 300 to 45~K.
No qualitative changes in the band are seen with further decreasing temperature below the transition. 

The second effect is the appearance of another band below 200~\cm-1.
The 0-200~\cm-1 energy region is important because this is where many new collective excitations are seen below T$_{c} = 34$~K as a result of lattice distortions and the opening of the spin gap.
By now it is established that below T$_{c}$ the in-plane ordering of the spin/charges on the outer Vanadium $d$ shell involves, in each ladder, a zig-zag pattern like the one shown in Fig~\ref{f58}b \cite{GrenierPRB02JolyPRB03}.
The band at low energies can thus be understood as a precursor of the strong and sharp collective modes in the ordered phase, see Section 3.
It is important that this continuum exist in an incipient form at 300~K and can be clearly seen already at 200~K.
Similarly with the band peaked at 680~\cm-1, this low energy counterpart gains substantial spectral weight from the low energy side with cooling.
We consider the data in Fig.~\ref{f58}a as proof that the transition at 34~K is preceded almost up to 300~K by strong fluctuations of the low temperature order and suggest that the reason the critical temperature is so low is not because of intra-ladder dynamics, but it is rather due to the phasing of the zig-zag charge order between coplanar and inter-layer ladders.
This statement is supported by frequency and temperature dependent electron spin resonance (ESR) linewidth, see Ref.~\cite{NojiriJPSJ00}.
At 36~K the authors of this work estimate a characteristic frequency of about 700~GHz ($\approx 25$~\cm-1), for the precursor fluctuations.

We also note that the topology of the ladder at quarter filling and the charge pattern below T$_{c}$ suggest that other magnetic terms are important beside the nearest neighbor exchange $J_{1}$ shown in Figs.~\ref{f53}b and \ref{f58}b.
For instance, if one compares in Fig.~\ref{f58}b the low temperature exchange path between two V atoms sitting on two adjacent rungs (which is related to $J_{1}$) to the exchange path between next nearest neighbors (which is related $J_{2}$), one can argue that the difference may not necessarily be large.
On the contrary, the superexchange $J_{2}$, taking place on a straight line across one vanadium and two O$_{leg}$ atoms, could be comparable to $J_{1}$. 
A quantitative estimation of possible competing interactions would be very welcome and important for the interpretation of the magnetic excitation spectrum.
The statements about the importance of other magnetic terms is supported by measurements of the temperature dependent magnetic susceptibility and its comparison with the theoretical expectations within the model of 1D AF Heisenberg chains characterized only by nearest neighbor exchange, see Ref.~\cite{JohnstonPRB00}.
The authors of this work show that the measured susceptibility $\chi(T)$ above 34~K is not in good quantitative agreement with the predictions of $S = 1/2$ uniform Heisenberg chain with only nearest neighbor interaction.
The data can be reconciled with theory only if additional magnetic exchanges are included and/or the superexchange itself is temperature dependent.
In view of the discussion above, we believe that a significant temperature dependence of the effective magnetic interactions as well as of the relative importance of competing exchange terms could be understood if one invokes the increasing charge imbalance on each V-O-V rung with cooling.

\subsection{Interpretation of the Raman continuum in terms of  multi-spinon Raman scattering}

In discussing the nature of the Raman band we have to acknowledge the existence of a substantial body of prior spectroscopic (Raman and IR) work, see for instance Refs.~\cite{DamascelliPRL98,GolubchikPopovicPopova97-99,KonstantinovicPSS99,KonstantinovicJPCM99,FischerPRB99}, as well as concomitant or subsequent to our study \cite{KonstantinovicPRB01and02,PopovaPRB02}.
Interesting properties were observed in addition to what has been discussed in the previous section.
One of them is its dependence on the Na concentration in \naxvo crystals \cite{KonstantinovicPRB01and02}.
With a decrease of the Na concentration a pronounced weakening in the continuum intensity as well as a downshift in energy from 680 to about 480~\cm-1 has been observed as $x$ was varied from from 1.0 to 0.85 (and implicitly decreasing the number of rung electrons).
Another property is related to the presence of a broad continuum of excitations in the far IR part of the optical conductivity data, extending from very low frequencies (below about 100~\cm-1) almost up to the electronic excitations seen around 1~eV, when the electric field was parallel to the $a$-axis \cite{DamascelliPRL98}.
This continuum, whose energy scale resembles much the one seen in Raman spectra, was interpreted by the authors of Ref.~\cite{DamascelliPRL98} in terms of ``charged two-magnon'' excitations.
In order to insure an electric dipole coupling to these double spin flip excitation the authors had to introduce a rather artificial charge asymmetry between the legs of each ladder, represented by an energy difference for the single particle on-site energy depending if it sits on the left or right of the V-O-V rung.
Moreover, the model, which includes the hopping parameters $t_{\perp}$ and $t_{\parallel}$ together with Coulomb repulsion $U$ for double site occupancy, neglects the Coulomb interaction between nearest neighbor sites which seems to be an important ingredient for the analysis of charge dynamics of the quarter filled ladder \cite{MostovoySSC00andPRB02}.
While the energy scales certainly suggest a common origin for the Raman and IR continua, the interpretation of both of these features is a matter of debate.

The common interpretation of the Raman continuum is that it is related to Raman active transitions between crystal split $d$ levels of the V atoms \cite{KonstantinovicJPCM99,KonstantinovicPRB01and02}, while the interaction with the phonons resulting in the so called Fano shape is due to electron-phonon coupling.
This assignment is not however based on specific calculations.
On the contrary, its energy scale seems to be too low as band structure calculations \cite{SmolinskiPRL98} show that the splitting between vanadium $d$ bands involves energies at least of the order of 2500~\cm-1. 
The temperature dependence showed in Fig.~\ref{f58} remains also unexplained in this scenario.
In fact, it was pointed out in Ref.~\cite{FischerPRB99} that with cooling the broad band should recover the discrete nature of an excitation between well defined atomic electronic levels.
The band remains however broad to the lowest temperatures (around 5~-~10~K) measured.

Although the possibility that the Raman continuum has a magnetic origin was mentioned in literature, this idea has been ruled out, see Refs.~\cite{KonstantinovicJPCM99,FischerPRB99,KonstantinovicPRB01and02} and the main argument was that this excitation appears only in $(aa)$ polarization and it is not present in $(bb)$ polarization.
The latter geometry was considered the expected scattering configuration where a two-magnon type excitation should be observed if one takes into account the spin structure shown in Fig.~\ref{f53}b and also the fact that, according to the Fleury-Loudon Hamiltonian \cite{FleuryPR68,ShastryPRL90}, the $b$-axis is parallel to the dominant exchange path.
A second reason was that a two-magnon continuum peaked at 680~\cm-1 would yield a value of the nearest neighbor superexchange $J_{1}$ too large in comparison to estimations from magnetic susceptibility \cite{JohnstonPRB00} and inelastic neutron scattering \cite{GrenierPRL01}.
We will return to this issue later in this section.
For the moment we mention that it was demonstrated in Figs.~\ref{f56} and \ref{f57} that the continuum, although weaker, is present also in the configuration when the incoming and outgoing photons are parallel to the ladder legs.
Moreover, a strong argument in our view against $d$ to $d$ transitions is the presence of the second overtone of this excitation around 1320~cm-1, see Fig.~\ref{f55}, whose weaker intensity suggests that it results from second order scattering.
There is no intrinsic reason that the crystal field split bands are distributed in such a way that another set of Raman allowed transitions is to be found at exactly double the frequency of the main peak.
Finally, there is a counterpart of this Raman continuum in the spin-Peierls compound CuGeO$_{3}$ whose origin is generally agreed on to be in double spin-flip processes \cite{LoosdrechtPRL96}.
The different energy scales of the continua in CuGeO$_{3}$ and \navo can be understood in a simple way if one takes into account the ratio of the dominant magnetic exchange interactions in these two compounds. 

The above arguments allow us to propose that the origin of the Raman band peaked at 680~\cm-1 is magnetic and that it represents a two-magnon like continuum of excitations.
Moreover, we will show in the following that this assignment is compatible with its temperature dependence, resonance and polarization properties, which further strengthen the identification of the origin of this band as magnetic.
We will also discuss the possibility of reconciling the obtained value for the magnetic superexchange with the ones existent in the literature.
Given the fact that in the high temperature phase \navo can be mapped onto an array of weakly interacting $S = 1/2$ AF chains (see Fig.~\ref{f53}), we will discuss first their characteristic excitations within the Heisenberg model.

{\bf Excitations out of a quasi 1D $S = 1/2$ AF chain and the two spin-flip Raman continuum~-- }
The purpose in this part is to show that a broad band as observed experimentally is consistent with the theoretical predictions for spin-flip excitations in a 1D AF $S = 1/2$ chain.
Consider a system like in Fig.~\ref{f58}b and the Hamiltonian:
\begin{equation}
H = J_{1} \sum_{(i,j) = NN} {\bf S}_{i} \cdot {\bf S}_{j} \ + \ J_{2} \sum_{(i,j) = NNN} {\bf S}_{i} \cdot {\bf S}_{j}
\label{e51}
\end{equation}
where ${\bf S}_{i}$ and ${\bf S}_{j}$ represent spins on chain sites $i$ and $j$ respectively while NN and NNN stand for nearest neighbor ($j = i \pm 1$) and next nearest neighbors in the same ladder ($j = i \pm 2$).
$J_{1}$ and $J_{2} = \alpha J_{1}$ are the corresponding superexchange integrals and we consider the case where both of them are positive.
\begin{figure}[t]
\centerline{
\epsfig{figure=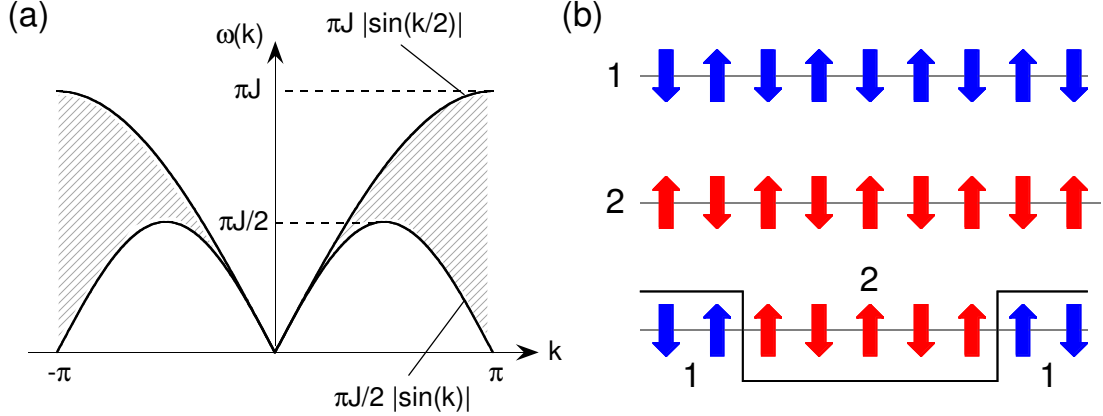,width=5.75in}
}
\caption{
(a) Excitation spectrum of a 1D $S = 1/2$ AF linear chain with nearest neighbor Heisenberg exchange $J$.
The lower line, $(\pi J / 2) | \sin(k) |$ is the dispersion for one spinon.
The $\pi J | \sin(k / 2) |$ curve marks the higher energy edge of the two-spinon continuum.
(b) Cartoon with a two spinon excitation.
$1$ and $2$ are two degenerate configurations in the AF ground state.
The two spinon excitation in the lower part can be thought of as a pair of domain walls between regions of type $1$ and $2$.
The total spin of this state compared to the states $1$ and $2$ amounts to one spin flip.
}
\label{f59}
\end{figure}
The NNN term leads to spin frustration, but due to the larger spatial separation in the high temperature phase it is generally expected that $J_{2}$ is only a fraction of $J_{1}$.
Let us look at the $J_{2} = 0$ case first. 
The excitations out of a $S = 1/2$ 1D AF chain in this limit are gapless domain wall solitons called spinons which have a dispersion given by $\omega(k) = \frac{\pi J}{2} | \sin(k) |$ \cite{FaddeevPL81}. 
Spinons carry a spin of 1/2 and they can be created only in pairs, each pair corresponding to one spin flip and thus having integer spin.
In some sense one can think about a magnon like excitation as being composed out of two spinons.
Similarly, a state with two spin flips (two-magnon Raman scattering) would correspond to the creation of four spinons.
Figure~\ref{f59}a shows the excitation spectrum of the spinons and the continuum of two-particle excitations.
Panel b is an intuitive visualization of a two-spinon excitation carrying a total spin $S~=~1$.

Here we come back again to a remark made in Chapter \cite{ChapterSCCO}: which way provides a better description of the spin dynamics, one that starts from true elementary excitations (in our case the spinons which are fractional spin excitations) or a description in terms of elementary triplet ($S = 1$) excitations, which are bosonic modes (corresponding to the known textbook magnons in the case where there is long range magnetic order and a semi classical approach applies)?
This question is probably most relevant for the $S = 1/2$ 1D AF chain with only NN interactions because this is the archetype of a gapless critical model whose true elementary excitations are known to be the spinons, and not triplons (elementary triplets) \cite{FaddeevPL81}.
This question was addressed by the authors of Ref.~\cite{SchmidtPRL03}, who perturbatively calculated spectral densities for various operators $R$ connecting the ground state to states with different number of excited triplets according to: $I(\omega) = \sum_{f} |<f| R |0>|^{2} \delta(\omega_{f} - \omega_{0} - \omega) = - \pi^{-1} Im [< 0 | R (\omega + \omega_{0} - H)^{-1} R | 0 >]$ where $H$ is given by Eq.~\ref{e51} with $J_{2} = 0$, $|f>$ and $|0>$ denote excited and the ground state while $\omega_{f}$ and $\omega_{0}$ represent their energies.
Using a continuous unitary transformation which conserves the number of elementary triplets and appropriate interactions $R$, they could evaluate separate relative contributions to $I_{tot} = \int_{0}^{\infty} d\omega \ I(\omega) = \sum_{n = 1}^{\infty} \ I_{n}$ from sectors with only one, two and three triplets (the total intensity being calculated using the sum rule $I_{tot} = <0|R^{2}|0> - <0|R|0>^{2}$).
The main result was that the sum rule is very well fulfilled by taking into account only the contributions up to (and including) three triplets \cite{SchmidtPRL03}, implying that there is no necessity to resort to fractional excitations.
It would be interesting to check if this statement remains true at all energies, an issue which could be addressed probably by evaluating relative contributions to energy and wavevector dependent spectral densities, $I(k,\omega)$. 

How does the excitation spectrum evolve with increasing the NNN frustrating parameter $\alpha = J_{2} / J_{1}$?
It turns out that for $0 \leq \alpha \leq \alpha_{c}$ the excitation spectrum remains gapless and qualitatively the same as in Fig.~\ref{f59} \cite{CastillaPRL95} and the true excitations are again spinons.
However, there is a critical value $\alpha_{c} \approx 0.24$ above which the low energy spectrum develops a finite spin gap.
The spin-spin correlations become short ranged from the power law fall-off characteristic of the uniform unfrustrated chain.
For an intuitive picture, the value $\alpha = 0.5$ is very interesting because in this case the ground state is doubly degenerate and it is known exactly to be made of products of NN singlets.
In the notation of Eq.~\ref{e51}, this means that for this value of $\alpha$ the two degenerate ground states are $|0>_{1} = 2^{N/2} \prod_{i = 1}^{N} ( |\uparrow_{2i-1}> |\downarrow_{2i}> - |\downarrow_{2i-1}> |\uparrow_{2i}> )$ and $|0>_{2} = 2^{N/2} \prod_{i = 1}^{N-1} ( |\uparrow_{2i}> |\downarrow_{2i+1}> - |\downarrow_{2i}> |\uparrow_{2i+1}> )$.
The value of $\alpha = 1/2$ for 1D chains is known as the Majumdar-Ghosh point \cite{MajumdarGhosh}.
This is also the moment to advertise the compound discussed in Chapter \cite{ChapterSCBO}, SrCu$_{2}$(BO$_{3}$)$_{2}$, which is the only experimental realization of a two-dimensional structure, well described by a NN and NNN Heisenberg terms, having an exactly known ground state.
Similarly to the 1D case, this ground state can be also written in terms of independent nearest neighbor singlet states.

As it is also discussed in the Chapter \cite{ChapterSCBO}, in the absence of spin-orbit coupling and because of the spin selection rules, Raman scattering usually couples to double spin flip states $via$ the photon induced spin exchange process \cite{FleuryPR68,ShastryPRL90}.
The effective spin Hamiltonian corresponding to this interaction in the off-resonance regime is given by:
\begin{equation}
H_{int} \propto \sum_{<i,j>} ({\bf e}_{in} \cdot {\bf r}_{ij}) ({\bf e}_{out} \cdot {\bf r}_{ij}) {\bf S}_{i} \cdot {\bf S}_{j}
\label{e52}
\end{equation}
where ${\bf r}_{ij}$ is the vector connecting these sites and ${\bf e}_{in}$/${\bf e}_{out}$ are the unit vectors corresponding to the incoming/outgoing polarizations.
Because the light wavevector is very small, the total momentum of the excitations probed has to be zero.
From Fig.~\ref{f59} one can infer that the energy range spanned by four spinon excitations with zero total momentum is quite large: from $\omega = 0$ up to $\omega  = 2 \pi J$.
One also expects that the Raman form factor is peaked in the middle of this region, at $\omega  = \pi J$, reflecting the divergence in the spinon density of states at this energy.
The large energy range corresponding to the multi-spinon continuum along with the well established inelastic light coupling to double spin flip excitations is thus compatible with the existence of the broad Raman continuum observed in Figs.~\ref{f55}, \ref{f56} and \ref{f58}.

\begin{figure}[t]
\centerline{
\epsfig{figure=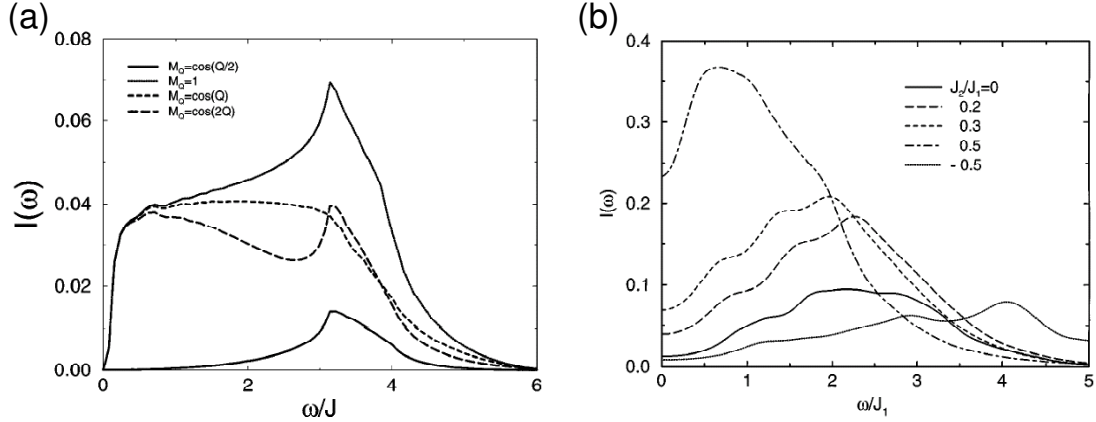,width=5.75in}
}
\caption{
(a) Raman intensity for four spinons evaluated with different matrix elements, $M = \cos(Q/2), 1, \cos(Q), \cos(2Q)$, from Ref.~\cite{SinghPRL96}.
A nearest neighbor (NN) Fleury-Loudon interaction (Eq.~\ref{e52}) corresponds to $M = \cos(Q)$ and a next nearest neighbor (NNN) interaction to $M = \cos(2Q)$.
(b) Raman intensity for the Heisenberg model (Eq.~\ref{e51}) for various ratios of the NNN exchange $J_{2}$ with respect to the NN term $J_{1}$.
Note that an increase of $J_{2} / J_{1}$ results in a downshift of the four spinon continuum spectral weight.
The results in this panel are also from Ref.~\cite{SinghPRL96}.
}
\label{f510}
\end{figure}

In fact numerical calculations using the Hamiltonian Eq.~\ref{e51} and an interaction of the form $H_{int} \propto \sum_{i} {\bf S}_{i} \cdot {\bf S}_{i + 2}$ revealed that the Raman intensity corresponding to the four spinon excitations corresponds indeed to broad continuum like feature, see Fig.~\ref{f510}a and Ref.~\cite{SinghPRL96}.
Note that this interaction is nothing else but a particular form of Eq.~\ref{e52} for $(bb)$ polarization (polarization along the chain direction) where a NNN term was chosen.
It is also important to observe that because of the commutation relations with the system Hamiltonian, the interaction $H_{int}$ could have been as well chosen to be $H_{int} = \sum_{i} {\bf S}_{i} \cdot {\bf S}_{i + 1}$, i.e. to describe spin exchange between NN spins, but in this case it would have been necessary to include NNN terms in Eq.~\ref{e51} so that $[H,H_{int}] \neq 0$.
While the exact shape of the four-spinon continuum from Fig.~\ref{f510}a does not exactly correspond to the experimental findings in \navo (Figs.~\ref{f55} and \ref{f56}) or CuGeO$_{3}$ (Ref.~\cite{LoosdrechtPRL96}), it nevertheless proves that the four spinon Raman continuum is an excitation relevant on an energy scale ranging from zero up to energies of about $6 J$ (which corresponds to the $0-2 \pi J$ range inferred from Fig.~\ref{f59}).
The overestimation of the low energy spectral weight has to do with the choice for the Raman matrix elements.
One can compare Fig.~\ref{f510}a with results obtained by exact diagonalization and/or mean field approximation for the frustrated spin chain \cite{MuthukumarPRB96,BrenigPRB97}.
These calculations reproduce better the experimental findings because the divergence at $\pi J$ is replaced by a broader peak and also because they eliminate the preponderant low frequency part. 

{\bf Discussion of the temperature dependence -- }
Here we intend to show that the temperature dependence from Fig.~\ref{f58}a is also consistent with an interpretation of the continuum in terms of multi-spinon excitations.
The experimental observation there was related to the softening of the Raman band with cooling from room temperature to about 40~K.
An interesting connection can be made between Fig.~\ref{f510}b and the discussion about the role of fluctuations and higher order magnetic exchange terms regarding the results in Fig.~\ref{f58}a-b.
In this respect, note also that, according to Eqs.~\ref{e51} and \ref{e52}, the inclusion of terms beyond the nearest neighbor (either in the system Hamiltonian or in the interaction Hamiltonians) are crucial for observing any inelastic magnetic signal.

Fig.~\ref{f510}b shows that there is a substantial overall downshift in the four-spinon spectral weight with increasing the NNN frustrating interaction $J_{2}$.
When we analyzed the data in Fig.~\ref{f58}a-b we brought evidence that the fluctuations of the low temperature order start at high temperatures and also that magnetic susceptibility results cannot be understood if only the NN exchange $J_{1}$ is taken into account.
In particular, we discussed that the tendency of the charges to arrange themselves in a zig-zag pattern can make the NNN exchange $J_{2}$ an important parameter.
So the reasoning goes as follows: (1) decreasing the temperature leads to a more and more pronounced charge disproportionation along the left and right legs of the ladder; (2) this may lead on one hand to a decrease of $J_{1}$ because the exchange path gets modified (the two electrons on a rectangular plaquette formed by two rungs will like to stay along the diagonal) and on the other hand to a relative increase of $J_{2}$ because the NNN Vanadium atoms become connected by a straight, shorter superexchange path; (3) according to Fig.~\ref{f510}, both effects in (2) will lead to a softening of the four spinon Raman continuum.
While still speculative (in the absence of a quantitative estimation of the charge imbalance as a function of temperature or of quantitative microscopic calculations of the relative strengths of the magnetic interactions) our point was to provide a possible basis for the understanding of the data in Fig.~\ref{f58}.

{\bf Discussion of polarization properties -- }
In a strictly 1D system, because any two spin sites can be connected only by vectors ${\bf r}_{ij} \parallel \hat{b}$-axis, the Fleury-Loudon polarization selection rules (Eq.~\ref{e52}) would not allow coupling in a polarization perpendicular to the chain direction, the $(aa)$ configuration.
However, we do observe scattering in this geometry.
 
Here is the place to reiterate some issues about the symmetry of the high temperature phase.
The importance of symmetry resides in the fact that it is the most direct way to establish the spin/charge pattern above T$_{c}$.
We discussed in the introduction that more recent X-ray scattering as well as NMR data support a centrosymmetric group in the high temperature phase and the existence of only one type of V atoms with an average valence of $+4.5$. 
This means that, because the rung hopping $t_{\perp}$ is the dominant term compared to the inter-ladder hopping $t_{xy}$ or the overlap along the chain direction $t_{\parallel}$ \cite{SmolinskiPRL98}, on each rung there is only one electron which is rapidly hopping from the left to the right side making (averaged over the timescale characteristic of each experimental probe) the two rung vanadium atoms look equivalent.
At a quantum mechanical level, one can think about the electronic wavefunction as a superposition of 'instantaneous' states having electrons arranged in different configurations so that when the charge density is averaged over the left and right legs of the ladder one gets (almost) equal fractional charges $q = e/2$.

In spite of the preference of the centrosymmetric group $Pmmn$ \cite{DamascelliThesis99,SmolinskiPRL98,MeetsmaACC98,LudeckePRL99} over the $P2_{1}mn$ group \cite{CarpyACB75}, this assignment is not entirely unambiguous and the analysis of the X-ray data becomes quite involved because both the crystallographic symmetries render almost identical diffraction patterns \cite{DamascelliThesis99}.
In fact a small, finite charge imbalance between V-O-V ladder legs existent even at room temperature could not be completely excluded.
This is fully consistent with the Raman data in Fig.~\ref{f58} which shows that the effects of the fluctuations of the low temperature zig-zag order can be tracked down at all measured temperatures above T$_{c} = 34$~K and that this temperature has to do mostly with the inter-ladder phase coherence.
This in turn means that locally and on relatively short time scales the charge pattern on each ladder is different from the average distribution seen by X-ray diffraction and that a cartoon for the high temperature phase like the one shown in Fig.~\ref{f53} is only an idealized approximation.

If one agrees with the notion that inside each ladder the charge/spin distribution is not strictly 1D, then the selection rules derived from Eq.~\ref{e52} allow scattering in $(aa)$ polarization.
So four spinon continuum should be seen in both $(aa)$ and $(bb)$ geometries, consistent with our experimental observations.
This also implies that the temperature dependence of the magnetic Raman band should track the evolution of the zig-zag order in the V-O planes, in particular we should observe an increase of its spectral weight with cooling, a process driven by closer proximity to the low temperature phase.
This is one aspect we already noted when discussing the Raman results, i.e. that the softening of the 680~\cm-1 peak is also accompanied by an increase in the spectral weight on the lower energy side.
Once the selection rules allow it, the peak in $(aa)$ can be imagined as being the result of the well known phonon induced spin exchange process: an electron-hole pair is created by an electron dipole interaction involving bonding-antiboding V $3d$ orbitals and O $2p$ bands, see Fig.~\ref{f54}; in the intermediate state two pairs of spinons are emitted before the collapse into an excited spin state.
Why is the signal in $(aa)$ stronger than in the expected geometry $(bb)$?
This has to do with the fact that the coupling when the electric field is perpendicular to the ladder is different from that along the $b$-axis and this is especially the case with the Raman vertex in resonance conditions.
We showed in Fig.~\ref{f56} that the intensity of the Raman band follows very closely the features seen in absorption, so one cannot talk about off-resonant conditions.
In this respect we note that quite different couplings in $(aa)$ with respect to $(bb)$ polarization of fully symmetric excitations are also observed for phonons, see Fig.~\ref{f55}.
It is known from the case of 2D cuprates that very strong enhancements of the magnetic signal can occur \cite{ChubukovPRL95} and they have been observed in resonant Raman spectra \cite{BlumbergPRB96}.
A quantitative understanding of the anisotropy in the $(aa)$ and $(bb)$ intensities would require specific evaluations of the Raman matrix elements, but in principle it is compatible with our observations

A zig-zag pattern also implies finite magnetic scattering in $(ab)$ polarization.
In Fig.~\ref{f57}d we show that the integrated intensity in this geometry is excitation energy dependent, but the 680~\cm-1 peak is not observed.
In fact this is not surprising, because the spin flip scattering in cross polarization should have a different form factor and it belongs to a different irreducible representation in an orthorhombic group.
As a result, it is a priori expected to have a different spectral shape than in parallel polarization.
Due to the distinct form factor in different scattering configurations, the Raman intensity integrates from different parts off the Brillouin zone and symmetry properties may lead to the cancellation of the zone boundary divergence in the density of states at $\pi J$.
Examples in this regard are again the 2D cuprates where the inclusion of terms beyond the nearest neighbors in the effective light coupling Hamiltonian led automatically to the appearance of magnetic scattering not only in B$_{1g}$, but also in A$_{1g}$ and B$_{2g}$ symmetry channels \cite{SinghPRL89}.
However, the spin pair scattering in each channel comes in with its own resonance and spectral properties.

{\bf Evaluation of the magnetic superexchange -- }
The peak position of the four spinon continuum at 680~\cm-1 (84~meV) (see Fig.~\ref{f55}) and the expected maximum in the Raman response at $\pi J_{1}$ reflecting the divergence in the density of states (see Fig.~\ref{f510}a) lead to an estimation for the NN superexchange $J_{1} \approx 27$~meV (218~cm-1 or 310~K).
Given the spectral weight downshift induced by the presence of the NNN term $J_{2}$ (Fig.~\ref{f510}b) we can say that this value is a lower bound for the NN superexchange.
Our value is in good agreement with the initial determination of $J_{1}$ from a fit to the magnetic susceptibility data \cite{IsobeJPSJ96} by the Bonner-Fisher prediction (Ref.~\cite{BonnerPR64}) as well as with the estimation from charged bi-magnon continuum observed in the optical conductivity data \cite{DamascelliPRL98}.
The interpretation of the IR absorption data relies on a global charge asymmetry between the ladder rungs.
Seemingly artificially introduced, this assumption may be regarded retrospectively in a more favorable light in the context of strong fluctuations of the low temperature order.
It is also possible that the difficulties in a precise distinction of the symmetry in the high temperature phase can be related to the same phenomenon. 
As for the magnetic susceptibility data from Ref.~\cite{IsobeJPSJ96}, it was pointed out in a more recent paper, Ref.~\cite{JohnstonPRB00}, that it could not be reproduced quantitatively in subsequent studies.
Moreover, the authors of this latter work obtained from an approximate fit (because they argued that the data cannot be quantitatively fit by a simple Bonner-Fisher expression involving only NN Heisenberg interactions) a value for $J_{1}$ which is almost twice as high as ours.

A fit to the experimental dispersion of the elementary triplet excitation in the low temperature phase obtained in a INS study, see Ref.~\cite{GrenierPRL01}, allowed the extraction of a value $J_{1} \approx 60$~meV, also about a factor of two higher than 27~meV.
Regarding this data we note that the maximum neutron energy transfer was 40~meV while the inferred maximum for the triplet dispersion was 93~meV.
Accordingly, the value of the energy at the zone boundary, which is of crucial interest to us (see Fig.~\ref{f59}), was not determined directly but only by using an approximate fitting function \cite{GrenierPRL01}.
Neither the authors of the INS work, nor those of Ref.~\cite{JohnstonPRB00} consider the possible influence of NNN exchange terms which we believe play an important role in determining the structure of the magnetic excitation spectrum.

We conclude by saying that our estimation for the lower bound of the superexchange, while not contradicting other analyses, is still too far apart to be explained by error bars.
In our view, a quantitative determination of the relevant $J's$ is still an open issue and the underlying reason is in the strongly fluctuating low temperature order and the induced frustration effects.
We also believe that the overall consistency between the interpretation of the strong Raman band in the high temperature phase in terms of multi-spinon excitations and our experimental findings strongly argues for its magnetic origin.
\newline

\section{Collective excitations in the low temperature \newline phase of \navo ($T~\leq~34$~K)}

\subsection{General features of the transition}

\begin{figure}[t]
\centerline{
\epsfig{figure=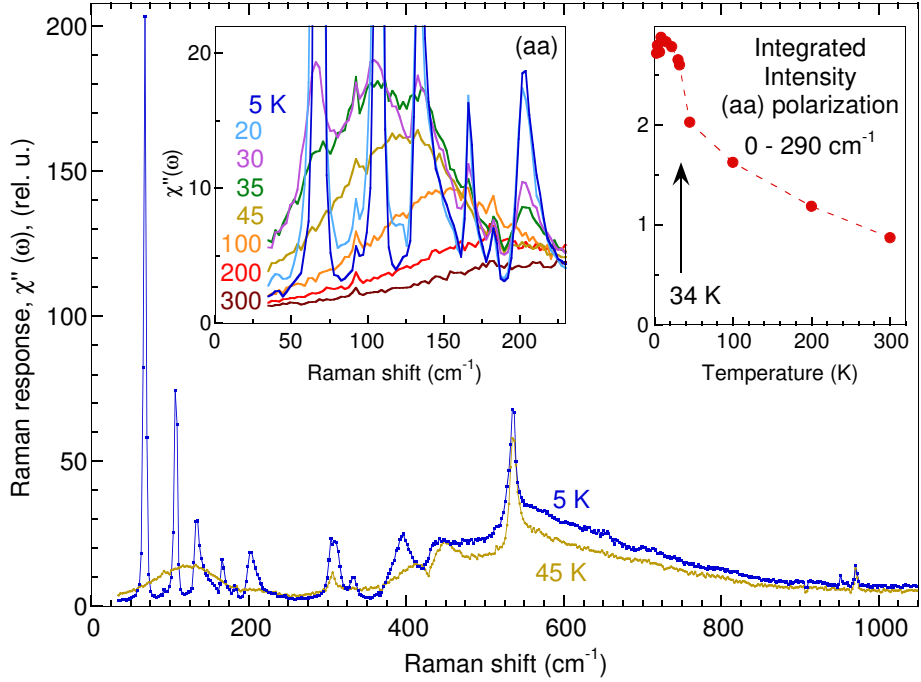,width=4.8in}
}
\caption{
The main panel shows the Raman response $\chi''(\omega)$ in $(aa)$ polarization for T~=~45 and 5~K.
The data is taken using the $\omega_{in} = 1.92$~eV excitation energy.
Left inset: Temperature dependence of the $(aa)$ polarized spectra in the 0-250 \cm-1 region.
The data for 45 and 5~K are zoomed in spectra from the main panel.
Right inset: Integrated intensity, $\int \chi''(\omega) \ d\omega$, between 0 and 290~\cm-1 as a function of temperature.
}
\label{f511}
\end{figure}

In the main panel of Fig.~\ref{f511} we show two Raman spectra taken at 45 and 5~K in $(aa)$ polarization.
One can notice that drastic changes occur across the transition at 34~K especially at low frequencies where many new modes appear.
The most prominent three features below 150~\cm-1 are a very strong mode at 66~\cm-1, another one around 105~\cm-1 and the sharp edge at 134~\cm-1 ($\approx 2 \cdot 66$~\cm-1) marking the onset of a continuum.
Other new excitations which are clearly seen in the T~=5~K spectrum are found around 166, 202, 308, 332, 393 and 949~\cm-1.
Besides these modes, other weaker excitations are seen around 122 and 182~\cm-1.
In Fig.~\ref{f58} we argued that the strong temperature dependence in the low frequency part of the spectra and the development of the low frequency continuum between 0 and 250~\cm-1 is a precursor feature signaling fluctuations of the low temperature order at temperatures above T$_{c}$.
That this is true can be seen in the left inset of Fig.~\ref{f511} which shows that below about 34~K, this broad excitation 'splits' into very sharp resonances.
While above T$_{c}$ the individual ladders do not know about each other at large distances, below 34~K a global phase coherence of the charge and lattice dynamics is established.
Also, the drop in the 0-200~\cm-1 continuum marks the opening of a gap in this energy region. 
The integrated Raman response in the 0-290~\cm-1 region, $\int \chi''(\omega) \ d\omega$ shown in the right inset of Fig.~\ref{f511}, increases with decreasing T almost all the way down to the lowest measured temperature, their weight relating most probably to the variation of the charge/spin order and lattice distortion in the low temperature phase.
\begin{figure}[t]
\centerline{
\epsfig{figure=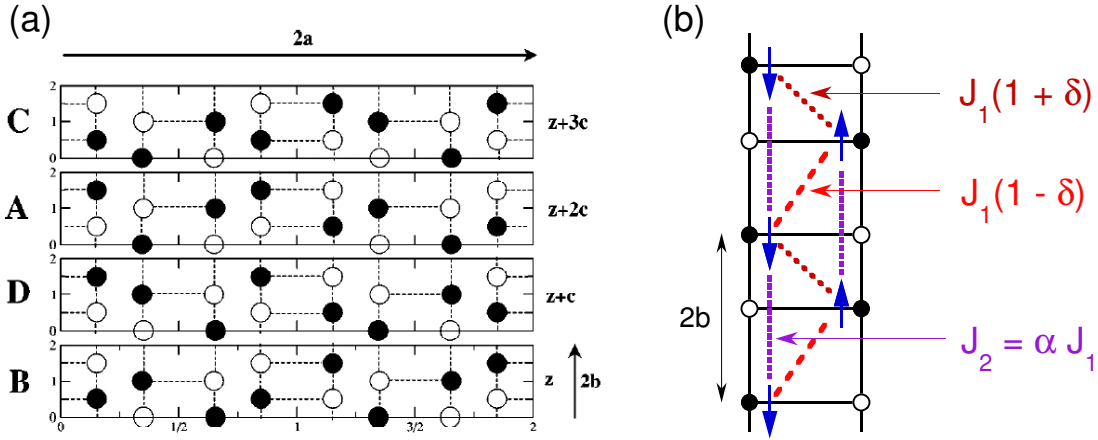,width=5.75in}
}
\caption{
(a) Schematic representation of the 3D charge/spin order (from Ref.~\cite{GrenierPRB02JolyPRB03}).
Black (open) circles are V$^{4+}$ (V$^{5+}$) atoms.
Each of the four rectangles shown represents a unit cell in the low temperature phase.
The sequence $B \ D \ A \ C$ represents one of the stacking sequences along the $c$-axis which are consistent with X-ray diffraction data, i.e. it has to the inter-layer phasing of the zig-zag order in each layer.
Note the doubling (along $a$ and $b$ directions) and the quadrupling of the unit cell along the $c$-axis.  
(b) Cartoon with the zig-zag order and relevant intra-ladder magnetic interactions.
The parameter $\delta$ characterizes the dimerization of the superexchange due to the unit cell doubling along the ladder legs.
Inter-ladder interactions are not shown.
}
\label{f512}
\end{figure}

How are these changes to be understood and what are the energy scales below T$_{c}$?
As for the crystal structure it is known that in the V-O layers there is a doubling of the unit cell along both the $a$ and $b$ directions \cite{FujiiJPSJ97,GrenierPRB02JolyPRB03,SmaalenPRB02}.
Moreover, in every ladder there is a zig-zag type ordering of the Vanadium valence electrons rendering, crudely speaking, V$^{4+}$ and V$^{5+}$ atoms on each rung of the ladder, see one rectangle in Fig.~\ref{f512}a.
Given the $2a \times 2b$ supercell, there are exactly four realizations of this superstructure in each plane, and they are denoted by A, B, C and D.
The existence of four such possibilities to accommodate the zig-zag order provided a clue for the understanding of the quadrupling of the lattice constant along the $c$-direction.
Initially an average face centered orthorhombic structure was proposed to explain the X-ray data at low temperatures \cite{SmaalenPRB02}.
The authors proposed a stacking disorder pattern along the $c$-axis rendering an effective global orthorhombic structure characterized by the face centered space group $Fmm2$.
However, this group has two mirror planes parallel to the $c$-axis which is at odds with the zig-zag order in each plane.
The resolution of this problem suggested in Ref.~\cite{SmaalenPRB02} was that because X-rays are sensitive only to the average structure, a model lacking mirror symmetry in each plane but having the appropriate stacking disorder along the $c$ axis can be in accordance with the measured diffraction pattern. 
A later resonant X-ray study argued for a lower (monoclinic) symmetry of the low temperature structure and proposed only two almost degenerate stacking patterns perpendicular to the ladder planes \cite{GrenierPRB02JolyPRB03}.
One of them, the B D A C model is shown in Fig~\ref{f512}a.
It was proposed that the true crystal structure below T$_{c}$ involves the existence of stacking faults of different possible patterns.
The data was suggested to reflect the presence of competing arrangements along the $c$ direction of nearly degenerate units.
This may lead to a rich phase diagram as a function of a parameter which could directly affect the inter-layer interactions.
Indeed, X-ray diffraction data under pressure, see Ref.~\cite{OhwadaPRL01}, show the development of a series of modulation wavevectors corresponding to an array of commensurate phases in the $P - T$ diagram.
The complicated observed sequences were qualitatively understood within the devil's-staircase-type phase transitions driven by two (presently unknown) competing interactions along $c$.

The above details and the discussion in the previous section give an idea about the difficulty in understanding quantitatively what exactly happens with the structure both above and below the transition.
However, what can be surely said about the changes in the in-plane magnetic interactions as a result of the transition?
A qualitative picture of what happens inside a given ladder can be obtained from Fig.~\ref{f512}b.
The unit cell doubling along the $b$ direction leads to an alternation in the NN superexchange $J_{1}$ which is quantified by the parameter $\delta$.
Due to the particular arrangement of charges in the zig-zag pattern, it seems that the NNN term may also play an important role in the spin dynamics.
The single ladder hamiltonian can be thus written as:
\begin{equation}
H = J_{1} \sum_{i} [(1 + (-1)^{i} \delta) {\bf S}_{i} \cdot {\bf S}_{i+1} \ + \ \alpha {\bf S}_{i} \cdot {\bf S}_{i+2} ]
\label{e53}
\end{equation}
The charge/spin order coupled to the lattice dimerization in the direction perpendicular to the ladders may lead to a complicated 2D pattern of magnetic exchanges (see for instance the discussion in the caption of Fig.~\ref{f514}).
\begin{figure}[t]
\centerline{
\epsfig{figure=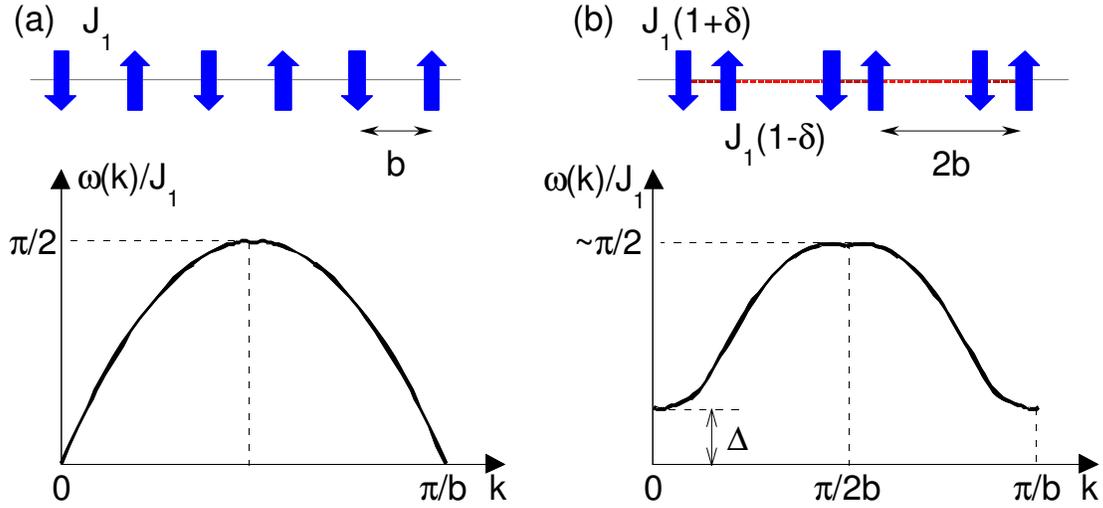,width=5.75in}
}
\caption{
(a) Cartoon showing the dispersion of the elementary excitations (spinons) in a uniform $S = 1/2$ AF chain (the same as in Fig.~\ref{f59}a).
(b) Qualitative changes in the (slightly) dimerized Heisenberg chain.
The unit cell along the chain becomes $2b$ and the alternation of superexchange is given the parameter $\delta$ (the same as in Fig.~\ref{f512}).
The elementary excitations are $S = 1$ triplets and a spin gap $\Delta$ opens up at the Brillouin zone center.
Apart from logarithmic corrections, $\Delta$ scales as $\delta^{2/3}$ \cite{TrebstThesis02}.
}
\label{f513}
\end{figure}
However, assuming a relatively weak inter-ladder coupling, which is confirmed experimentally by the small dispersion of the magnetic modes along $a$-axis, one can assume that the backbone Hamiltonian for the magnetic interactions is given by Eq.~\ref{e53}, i.e. a dimerized and frustrated 1D AF $S = 1/2$ chain.

We will discuss later the excitations from dimerized and frustrated chains in more detail.
This topic has been the focus of many theoretical studies \cite{TrebstThesis02,Dimerization,SchmidtPRB04}.
Here we are preoccupied only with the main result of the dimerization of the NN exchange, which is shown in Fig.~\ref{f513}a-b.
If in the uniform case, panel a, the excitation spectrum is gapless (see also Fig.~\ref{f59}), any finite dimerization $\delta$ will open up a spin gap at the Brillouin zone center.
This gap, denoted by $\Delta$, scales as $\Delta \propto \delta^{3/2}$ apart from logarithmic corrections \cite{TrebstThesis02,Dimerization}.
Given the fact that in the absence of any dimerization a gap can be opened only due to frustration for $\alpha \geq \alpha_{c} \approx 0.24$, one can see that in the $(\delta, \alpha)$ parameter space the spin gap and the magnetic excitation spectrum in general can be quite complicated.
Results of calculations for concrete sets of parameters are shown for instance in Fig.~\ref{f517}.
This figure will be discussed later in connection with the possible observation of Raman active magnetic collective modes in the low temperature phase.
As for the short wavelength excitations, we mention that in general, for quite a wide range of parameters $\delta$ and $\alpha$, the energies of the elementary triplet excitations at the zone boundary are not strongly renormalized from $\pi J_{1} / 2$, which is the value corresponding to the uniform NN AF spin chain. 

The low temperature phase has been quantitatively explored by various techniques.
INS data show that below T$_{c}$ there are two close-by gapped magnon excitations which have a large dispersion along the $b$-axis (inferred to be of about 80~meV - 645~\cm-1) and the corresponding $k = 0$ values between 8 and 11~meV (see the caption of Fig.~\ref{f514} for a discussion) \cite{GrenierPRL01}.
The magnitude of the $a$-axis modulation is much smaller, of the order of only 1~meV, confirming weak inter-ladder interaction.
We note that one of the spin gaps found by INS is situated at 8.3 meV (66.9~\cm-1), an energy which, within error bars, is equal to the one corresponding to the very strong Raman active mode seen at 65.9~\cm-1 in Fig.~\ref{f511} below T$_{c}$.
Note that the double folding of the branch corresponding to the 8.3~meV spin gap (shown in Fig.~\ref{f514}b) leads to the appearance of another Brillouin zone center feature with an energy of 10.9~meV (87.9~\cm-1).
The presence of spin-orbit coupling allows the observation of some of these elementary triplet excitations also in far IR \cite{RoomPRB04} and ESR spectra \cite{LutherJPSJ98}.
A low temperature gap of 8.13~meV (65.5~\cm-1) is determined with high resolution by these two techniques.

We observed so far that the transition at 34~K involves several aspects: (1) crystallographic distortions leading to doubling of the in-plane lattice  constants along and across the ladder legs as well as a quadrupling of the unit cell in a direction parallel to the $c$-axis; (2) in-plane zig-zag charge ordering which has a very complicated pattern in a direction perpendicular to the plane; (3) development of gapped magnetic branches (with spin gaps in the 9~meV energy range) which are much more dispersive along the ladder legs than in the transverse direction.
The question is what is the driving force of this transition?
Is it a spin-Peierls transition as it was initially thought, is it driven only by Coulomb forces which stabilize the zig-zag ordering, leading also to lattice distortions $via$ electron-phonon coupling or does it occur mainly as a result of a structural instability?
\begin{figure}[t]
\centerline{
\epsfig{figure=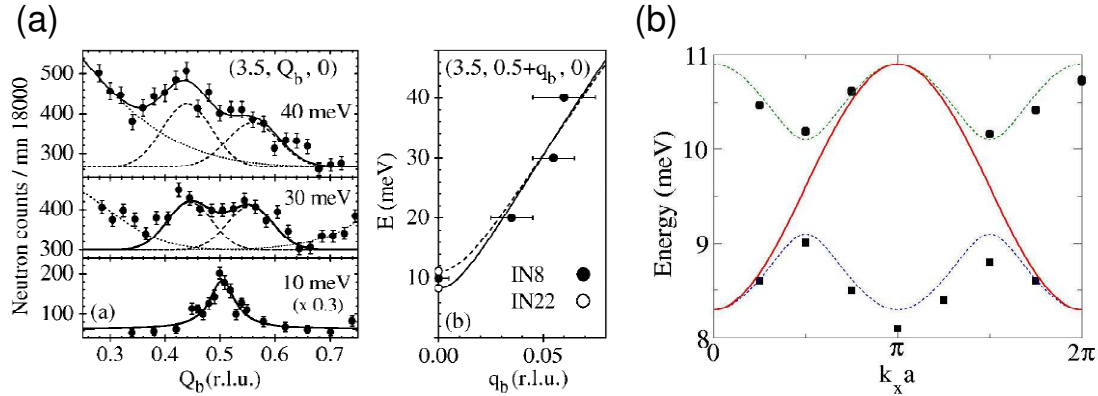,width=5.75in}
}
\caption{
Inelastic neutron scattering data from Ref.~\cite{GrenierPRL01}
(a) Left panel: Constant energy scans along the chain direction.
The wavevector $Q_{b}$ is defined in units of $1/b$.
Right panel: Energy dispersion of the elementary triplet excitation along the chain direction ($b$-axis); $q_{b} = Q_{b} - 0.5$.
The empty and filled circles at $q_{b} = 0$ in this figure correspond to data taken on two thermal neutron spectrometers, IN8 and IN22.
(b) Dispersion along the $a$-axis for $Q_{b} = 1$ of the elementary triplet branch showing the folding of this excitation due to the unit cell doubling in the direction perpendicular to the ladders.
Consequently, its periodicity is $\pi / a$.
The red line corresponds to the unfolded branch; note that its periodicity is $2 \pi / a$.
According to the interpretation in Ref.~\cite{GrenierPRL01}, in the Brillouin zone of the high temperature phase there is a second branch analogous to the one shown by the red line and very close in energy, whose dispersion is phase shifted by $\pi / 2$.
The reason this branch appears is because the superexchange $J_{1}$ are slightly different in adjacent ladders as a result of charge/spin ordering.
In the low temperature phase this second branch gets folded too (see Fig.~4 in Ref.~\cite{GrenierPRL01}), and the data in (a) refers to this particular excitation.
In the end, there are four Brillouin zone center modes corresponding to 8.3 (66.9) and 10.9 (87.9)~meV~(\cm-1), from the first branch shown in panel (b) and to 9.1 (73.4) and 10.1 (81.5) eV (\cm-1) from the second branch. 
}
\label{f514}
\end{figure}
Note that in the latter two scenarios the opening of a spin gap in the magnetic excitation spectrum is a by-product of the lattice/charge ordering rather than the main cause of the transition.

The arguments against an ordinary spin-Peierls scenario were mentioned in the introduction and they rely mainly on the absence of magnetic field effects on T$_{c}$, the absence of other incommensurate states at high fields and the observation of a large entropy release at the transition which could not be accounted by the spin degrees of freedom alone.
Seo and Fukuyama proposed in Ref.~\cite{SeoJPSJ98} a model Hamiltonian containing single $d_{xy}$ V orbitals and taking into account on-rung, ladder leg and inter-ladder hopping parameters $t_{\perp}$, $t_{\parallel}$ and $t_{xy}$ (see Fig.~\ref{f53}), as well as on site ($U$) and nearest neighbor Coulomb interactions $V_{\perp}$, $V_{\parallel}$ and $V_{xy}$ (these latter notations are in correspondence with the notations used for the hopping parameters).
Three stable AF solutions were found within the Hartree approximation and they correspond to: (1) the uniform phase (all V atoms are equivalent, i.e. what is thought to be realized in the high temperature phase); (2) the chain type having V$^{4+}$ atoms on a given ladder leg and V$^{5+}$ ions on the other (which would correspond to the initial room temperature crystal structure determination \cite{CarpyACB75}); (3) the zig-zag type, which is the one shown in Fig.~\ref{f512}.
It was found that for a large (and relevant) range of parameters, including the ones proposed in Ref.~\cite{SmolinskiPRL98}, the lowest energy state (above a small critical value of the intersite Coulomb interaction defined as $V = V_{\perp} = V_{\parallel} = V_{xy} / \sqrt{2}$) is one having zig-zag order with alternate charge disproportionation on each rung.
Depending on the relative magnitude of the hopping parameters it was proposed that the drop in susceptibility below T$_{c}$ is related to a singlet state formation of essentially localized pairs of spins.
The zig-zag order was also found consistent with the observed in-plane unit cell doubling. 
As a result, the authors suggested that the charge ordering due to Coulomb interaction is the driving force of the transition at 34~K in \navo \cite{SeoJPSJ98}.

Using a similar Hamiltonian except for excluding double occupancy of V $d_{xy}$ orbitals, equivalent to making $U \rightarrow \infty$, a subsequent study confirmed the stabilization of the zig-zag pattern \cite{MostovoySSC00andPRB02}.
The authors of Ref.~\cite{MostovoySSC00andPRB02} also pointed out that this kind of charge ordering along with inter-ladder correlations will automatically lead to an alternation of the exchange integrals along the ladder direction, $J_{i,i+1} = J_{1} (1 + (-1)^{i} \delta)$, where $i$ is a rung index.
This alternation will open up a spin gap and lead to an exponential drop of the susceptibility below T$_{c}$.
Another important point was related to the interaction between charges $via$ lattice distortions: the zig-zag structure is also favored by the displacements of the rung O atoms which are pushed by the large $V^{4+}$ ions.

More recent exact diagonalization results, see Ref.~\cite{AichhornPRB04}, also support the prominent role of the Coulomb interactions for the 34~K transition in \navo.
The authors highlight the importance of phonons in the stabilization of the zig-zag structure, especially of the alternate $c$-axis displacements of the V atoms in the low temperature phase, which were seen in X-ray data \cite{LudeckePRL99}.
It was found that the coupling of the electrons to the lattice can substantially reduce the critical value of the intersite Coulomb repulsion (see the previous paragraph) at which the formation of the charge ordered state occurs.

Interestingly, besides the low energy spin excitations arising as a result of dimerization, the characteristic low energy charge excitations are proposed to be soliton-like modes, similar to some degree to the spinons which are domain walls between degenerate ground state configurations, see Fig.~\ref{f59}.
In this case, for an individual ladder there are two equivalent zig-zag patterns forming the ground state, the difference between them being a lattice translation with a high temperature lattice constant $b$.
A low energy kink-like excitation will involve local on-rung electron hopping and will smoothly interpolate between the two degenerate configurations, see for instance Fig.~6 in Ref.~\cite{AichhornPRB04}.
One can conclude this section dedicated to the nature of the low temperature phase by saying that both experimental and theoretical works are strongly in favor of a transition driven by Coulomb interactions, with the spin gap formation playing only a secondary role.

\subsection{Observation of a folded triplet excitation. Selection rules and interpretation}

\begin{figure}[t]
\centerline{
\epsfig{figure=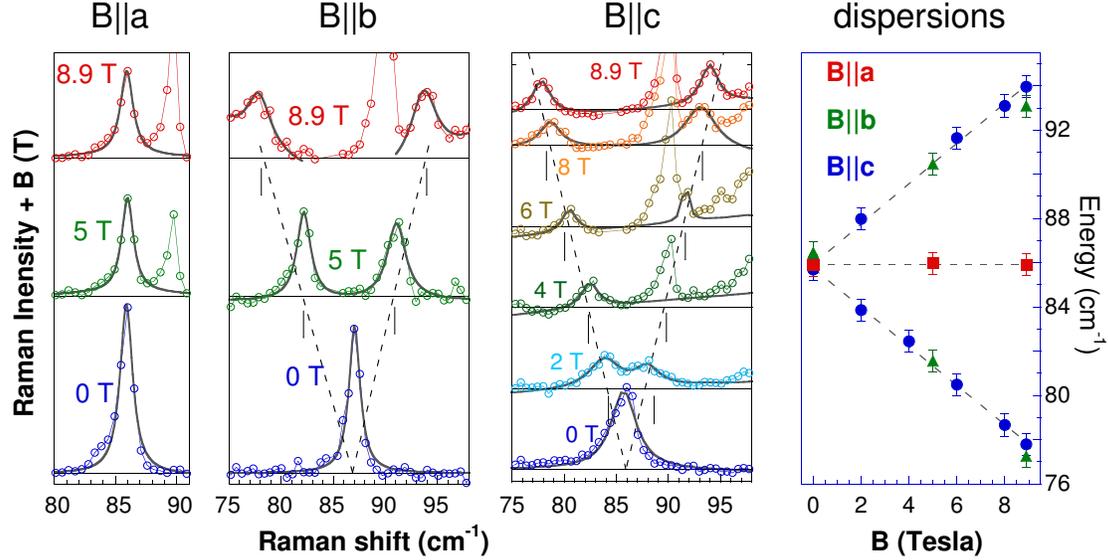,width=5.75in}
}
\caption{
From left to right, the first three panels show Raman response at 5~K as a function of magnetic field.
Each spectrum is off-set by an amount equal to the value of the field at which is was taken.
The colored symbols are data and the thick dark grey lines through the data points are results of Lorentzian fits to the triplet component(s).  
The feature appearing around 90~\cm-1 in finite fields is a phonon which exists also in the high temperature phase.
Note that the 86~\cm-1 mode splits for magnetic fields ${\bf B} \parallel \hat{b}$ and ${\bf B} \parallel \hat{c}$ but it does not shift for ${\bf B} \parallel \hat{a}$-axis.
Panel (d) shows the dispersion with field of the 86~\cm-1 mode for the three field configurations (squares for ${\bf B} \parallel \hat{a}$; triangles for ${\bf B} \parallel \hat{b}$ and circles for ${\bf B} \parallel \hat{c}$).
}
\label{f515}
\end{figure}

{\bf Polarization properties and selection rules in magnetic fields -- }
We discussed in the previous section that INS data reported the existence of Brillouin zone center spin flip excitations around 11~meV.
At low temperatures and in $(ab)$ polarization we observe a relatively weak feature around 86~\cm-1, see Fig.~\ref{f515}.
What relatively weak means compared to other features seen in cross and parallel polarizations can be inferred from in Fig.~\ref{f518} where this feature is denoted by $T$ in the lower left panel.
In finite external fields this feature has the following properties: it shows no shift for ${\bf B}$ parallel to the $a$-axis but we observe a splitting for the other two orthogonal directions of the external magnetic field.
Both the upward and downward dispersing branches change their energies in an approximately linear fashion, with a proportionality factor given by one Bohr magneton $\mu_{B}$.
This shows that we are looking at a $S = 1$ excitation and we identify it with the elementary zone center triplet seen by neutron scattering.
Note that in finite fields there is an extra mode around 90~\cm-1.
This is a fully symmetric phononic feature which appears because in finite magnetic fields, the orthorhombic (or even smaller) crystal symmetry leads to a Farady rotation of the polarization inside the sample preventing the observation of clean selection rules. 
In fact the finite field spectra were taken  after the polarization optics were rotated in order to minimize the intensity of strong features seen in parallel polarization (for instance the 66~\cm-1 mode from Figs.~\ref{f511} and \ref{f518}).

Fig.~\ref{f515} reveals an interesting behavior of the 86~\cm-1 mode.
Because of the spin selection rules, Raman scattering usually couples to singlet ($S = 0$) excitations.
The observed magnetic field dependencies clearly indicate the triplet nature of this mode.
This is possible in the presence of spin-orbit coupling which mixes the spin and orbital degrees of freedom, see the Introduction for a more detailed discussion.
In what follows we try to identify the coupling mechanism responsible for the experimental observations from Fig.~\ref{f515} as well as the corresponding selection rules.

{\bf The role of the antisymmetric Dzyaloshinskii-Moriya interaction -- }
It is often possible to capture the effects of certain spin-orbit interactions by writing effective spin Hamiltonians.
Such a Hamiltonian was inferred from macroscopic considerations in 1958 by Dzyaloshinskii \cite{DzyaloshinskiiJPCS58} and two years later by Moriya \cite{MoriyaPR60}, who derived a similar expression from a microscopic approach which needed superexchange (wavefunction overlap) and spin-orbit coupling as the only ingredients.
If one writes down the most general bilinear interaction between two spins at lattice sites $i$ and $j$ in the form $\sum_{\alpha,\beta} \ S_{i}^{\alpha} J_{ij}^{\alpha \beta} S_{i}^{\beta}$ and considers only the antisymmetric part of the $J_{ij}^{\alpha \beta}$ tensor, one obtains the general form of the Dzyaloshinskii-Moriya (DM) interaction:
\begin{equation}
H_{DM} = \sum_{(ij)} \ {\bf D}_{ij} \cdot ({\bf S}_{i} \times {\bf S}_{j})
\label{e54}
\end{equation}
This type of interaction appears usually in crystals with lower symmetry and it is often responsible for the phenomenon of weak ferromagnetism.
An example in this sense is the La$_{2-x}$Sr$_{x}$CuO$_{4}$ system discussed in more detail in \cite{GozarPRL04}.

The importance of the DM interaction was already suggested by ESR experiments which detected a considerable anisotropy of the absorption lines with respect to the magnetic field orientation \cite{LutherJPSJ98,NojiriJPSJ00}.
The direct observation of singlet-triplet transitions at low temperatures in ESR \cite{LutherJPSJ98} and low frequency IR absorption spectra \cite{RoomPRB04} also proposed that the DM interaction is the mechanism which allows the coupling of the photon field to single magnon excitations.
We note that below T$_{c} = 34$~K, the local crystal symmetry is low enough to allow the presence of DM type terms between neighboring spins.
In a theoretical study, Valenti \emph{et al.} studied Raman scattering in quasi-1D AF spin chains.
The main conclusion of this paper is that a single magnon excitation probed by Raman scattering has a very clear experimental signature: it should show no splitting in an external magnetic field parallel to ${\bf D}$ and it should split in two branches for a field perpendicular to the DM vector.

\begin{figure}[t]
\centerline{
\epsfig{figure=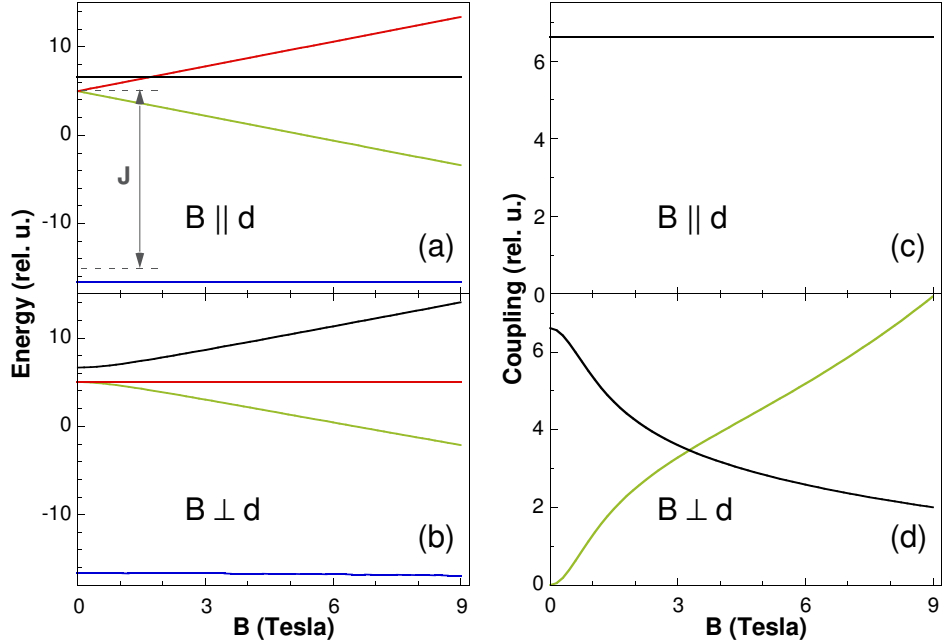,width=4.9in}
}
\caption{
Energy levels and light couplings as a function of magnetic field for a system of two spins with Dzyaloshinskii-Moriya interaction.
The Hamiltonian reads $J {\bf S}_{1} \cdot {\bf S}_{2} + {\bf D} \cdot ({\bf S}_{1} \times {\bf S}_{2}) + g \mu_{B} {\bf B} \cdot ({\bf S}_{1} + {\bf S}_{2})$ with ${\bf D}$ chosen parallel to the $\hat{z}$ axis.
The way the energy levels split for ${\bf B} \parallel {\bf D}$, panel (a), and for ${\bf B} \perp {\bf D}$, panel (b), are illustrated for $J = 20$ and $D = 12$.
Panels (c) and (d) show the moduli square of the nonvanishing matrix elements of the ${\bf S}_{1} \cdot {\bf S}_{2}$ operator between the ground state and the three excited states (the same color coding is used as on the left side of the figure).
}
\label{f516}
\end{figure}

The authors of Ref.~\cite{ValentiPRB00} also noted that there is no experimental confirmation of their prediction and so it is up to now to the best of our knowledge.
Inspection of Fig.~\ref{f515} shows that what we observe  experimentally resembles closely the theoretical predictions if ${\bf D} \parallel \hat{a}$.
Although the Raman selection rules for accessing the elementary triplet states may look complicated and were derived in Ref.~\cite{ValentiPRB00} based on a non-trivial cluster-model approach, we believe that some understanding can be gained if we look at the problem from the weakly coupled dimer limit.
That is, let us assume that the ladders are magnetically decoupled and in each individual ladder the alternation parameter $\delta$ from Eq.~\ref{e53} is very close to unity.
In this limit one deals in the zero order approximation with independent antiferromagnetically coupled spin dimers.
Let us include an intra-dimer DM interaction of the form given by Eq.~\ref{e54} and diagonalize this two spin problem in the presence of an external magnetic field.
The results, exemplified for $J = 20$ and $D = 0.6 J$ (which are values not extracted from any experimental data), are shown in Fig.~\ref{f516}.
This simple exercise is also useful for the understanding of some of the results in the \scbo chapter \cite{ChapterSCBO}, where a slightly more complicated system (four spins instead of two) is used in order to understand the low energy magnetism in the quantum spin system the SrCu$_{2}$(BO$_{3}$)$_{2}$ orthoborate.
However, in spite of the additional complication, many of the features discussed here are more visible and apply to the orthoborate case.

Panels (a) and (b) show the energy levels as a function of magnetic field for two directions: perpendicular and parallel to the DM vector.
In the absence of field and for ${\bf D} \equiv 0$ we have a singlet at energy $-3 J / 4$ and a triple degenerate $S = 1$ excitation at $J / 4$.
A DM interaction (chosen parallel to the $z$-axis) in zero field will mix the singlet and the triplet states, in particular the vectors corresponding to zero component of the $z$ projection of the total spin operator $S^{z} = S_{1}^{z} + S_{2}^{z}$.
As a result the states with $S^{z} = \pm 1$ will remain at $J$ and the $1/\sqrt{2} (|\uparrow>_{1} |\downarrow>_{2} \mp |\downarrow>_{1} |\uparrow>_{2})$ states (corresponding to the singlet ground state and excited $|S = 1, S^{z} = 0> \equiv |1,0>$ state respectively) will repel each other.
The $|0,0>$ ground state will have an energy $\frac{-3J}{4} + \frac{J}{2} \left ( 1 - \sqrt{1 + \frac{D^{2}}{J^{2}}} \right ) \approx \frac{-3J}{4} - \frac{D^{2}}{4J}$ for $D \ll J$  while the $|1,0>$ state will be at $\frac{J}{4} + \frac{J}{2} \left ( 1 + \sqrt{1 + \frac{D^{2}}{J^{2}}} \right ) \approx \frac{J}{4} + \frac{D^{2}}{4J}$ for $D \ll J$.
If the field is parallel to the DM vector, the $S^{z} = \pm 1$ states will split with one Bohr magneton per Tesla while the other two states will not change their energies.
If the field is perpendicular to the DM vector the mixing of the states will be different and this will lead to a different splitting of the triplet multiplet, see Fig.~\ref{f516}b.
Although at high fields the dispersion with field is again linear, the derivative of the energy with respect to the field as $B \rightarrow 0$ is vanishing in this second case.
The curvature of the dispersion at small fields is a measure of the strength of $D$.

Having in mind that the Fleury-Loudon coupling contains sums of scalar products of spins, see Eq.~\ref{e52}, in Fig.~\ref{f516}c-d we evaluated the moduli square of the matrix elements of the ${\bf S}_{1} \cdot {\bf S}_{2}$ operator and in these two panels we show all the non-vanishing terms for both field configurations.
In zero field the coupling is finite only for the $|1,0>$ state.
If ${\bf B} \parallel \hat{z}$ no additional coupling appears and the intensity of the $|0,0>$ $\leftrightarrow$ $|1,0>$ transition is field independent.
If ${\bf B} \perp \hat{z}$ there will be a finite coupling to the up and down dispersing branches.
The crossing points of intensities in panel (d) is also a measure of $D$.
Overall, what these two panels say is that if the magnetic field is parallel to the DM vector one should see only one mode which does not change its energy with field, while if it is perpendicular to it two modes should be seen in finite fields, and both of them should shift with increasing field.
The conclusions of this simple model are essentially the same as those of Ref.~\cite{ValentiPRB00} which says that a local picture in terms of independent dimers is qualitatively appropriate.

According to the experimental results from Fig.~\ref{f515}a-d and the theoretical predictions from Ref.~\cite{ValentiPRB00} our data can be understood as a result of the Raman coupling to one magnon excitations through the DM interaction with ${\bf D} \parallel \hat{a}$-axis in the low temperature phase.
A quantitative analysis of the intensity dependence is not possible because of the induced changes in the polarization directions inside the sample when magnetic fields are applied (changes that could not be controlled rigorously) as well as because of the appearance of the 90~\cm-1 feature which obscured the upward dispersing branch.
From the observed linear (within our resolution) magnetic field dependence of the modes energies as well as from the similar spectral weights of the split triplet components at all measured fields in the ${\bf B} \parallel \hat{b}$ and/or ${\bf B} \parallel \hat{c}$ configurations ($2 \leq B \leq 8.9$) we infer that possible non-linearities in the energy dispersions or strong variations in intensities take place below the value $B = 2$~T. 
Clearly, the intensity variations predicted in Fig.~\ref{f516}d are not seen in the experimental data which shows that the two peaks in panels (b) and (c) have similar intensities.
We think that this discrepancy occurs because the actual spin structure is much more complicated than a simple ensemble of uncoupled spin dimers and higher order spin interactions and possibly inter-ladder interactions have to be taken into account.

We clarified the existence of a single magnon excitation at 86~\cm-1 in Raman scattering spectra.
The feature is weak compared to other observed excitations in the Raman data.
The question we want to address in the following is whether we observe other magnetic modes, in the singlet channel, besides phononic excitations.
This is the topic of the following section.

\subsection{Do we observe magnetic bound states below the two-particle continuum?}

{\bf Excitations out of 1D $S = 1/2$ AF chains with dimerization and frustration -- }
We argued above that the relevant spin Hamiltonian for an individual ladder is given by Eq.~\ref{e53} so a closer look at the magnetic excitation spectrum of a dimerized and frustrated spin chain is useful.
We discussed there that any finite dimerization parameter $\delta$ and/or values of the frustration parameter $\alpha$ above a critical value will open up a spin gap $\Delta = \Delta(\delta,\alpha)$.
Qualitatively, the dispersion of the elementary triplet (one magnon) excitation given by $\epsilon(k)$ looks like in Fig.~\ref{f513}b.
The two triplet continuum (defined as the range of energies $\omega (k) = \epsilon(k_{1}) + \epsilon(k_{1})$ with the wavevectors satisfying $k = k_{1} + k_{2}$)  will have a minimum at $k = 0$ and starts from an energy $\epsilon = 2 \Delta$.
It turns out that the Hamiltonian of Eq.~\ref{e53} also allows for the existence of two triplet bound states, which means that at certain wavevectors there are two triplet excitations with energies below the continuum \cite{TrebstThesis02,Dimerization,SchmidtPRB04}.

\begin{figure}[t]
\centerline{
\epsfig{figure=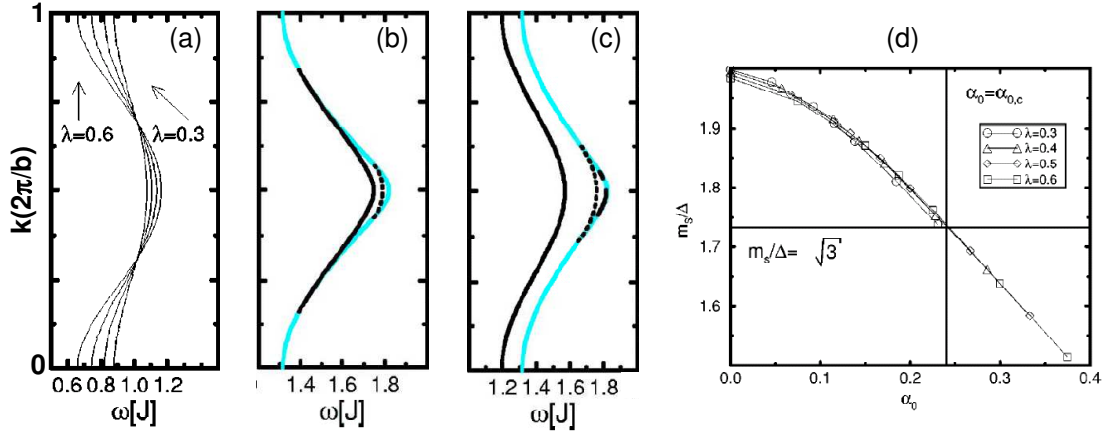,width=5.75in}
}
\caption{
Results of perturbative calculations using continuous unitary transformations for the excitations of a 1D AF chain with dimerization and frustration (from Ref.~\cite{SchmidtPRB04}).
The results shown here are for the NNN exchange $J_{2} = \alpha J_{1}$ with $\alpha = 0.25 (1 - \delta)$ and the energies are given in units of $J_{1} (1 + \delta)$.
The parameter $\lambda$ is defined in terms of the dimerization $\delta$ as $\lambda = (1 - \delta) / (1 + \delta)$.
(a) Elementary triplet dispersion $\omega(k)$ for $\lambda = \{0.3, 0.4, 0.5, 0.6\}$ which translates into $(\delta,\alpha) = \{(0.54,0.115); (0.43,0.143); (0.33,0.167); (0.25,0.187)\}$.
Panels (b) and (c) show two-triplet bound states (black lines) with total spin $S = 1$ and $S = 0$ respectively calculated for $\lambda = 0.6$ ($\delta = 0.25$).
The lower bound of the two particle continuum is shown by the light blue line.
(d) The ratio of the singlet ($S = 0$) bound states (at $k = 0$) to the value of the spin gap as a function of $\alpha$ (denoted by the authors of Ref.~\cite{SchmidtPRB04} by $\alpha_{0}$) for several dimerization parameters.
The vertical line corresponds to the critical frustration where a gap opens in the spin excitation spectrum in the absence of any dimerization (see the discussion related to Fig.~\ref{f59}) and the horizontal line denotes $\sqrt{3}$.
}
\label{f517}
\end{figure}

An example of a perturbational calculation \cite{SchmidtPRB04} of single and two triplet dispersions for certain points in the $(\delta,\alpha)$ parameter space is shown in Fig.~\ref{f517}.
Note that in these figures the wavevectors are on the vertical scales and the energies are on the horizontal axes.
Panel (a) shows elementary triplet excitations for several points in the $(\delta,\alpha)$ parameter space.
The dispersion is similar to Fig.~\ref{f513}b, there is a rapid variation of the zone center values but not much happens at the Brillouin zone center where the energies are not far from $\pi J_{1} / 2$ (note that the unit of energy in this figure is not $J_{1}$ but $J_{1} (1 + \delta)$).
Results for the two-particle sectors are shown in panels (b) and (c) and one can see the continuum marked by the light blue color.
One important thing to note in these two figures is the occurrence of two-triplet bound states whose dispersions are shown by black (solid, dashed or dash-dotted lines) lines.
Panel (b) shows the results for states in which two triplets are bound into a $S = 1$ state while panel (c) shows the same thing but for triplets bound in a state having a total spin $S = 0$.
For each case there are more than one bound states, the difference for the numerical parameters shown being that the binding energies of the triplet bound states are finite only at finite values of $k$ while for the singlet channel there is a branch split from the continuum in the whole Brillouin zone.
\begin{figure}[t]
\centerline{
\epsfig{figure=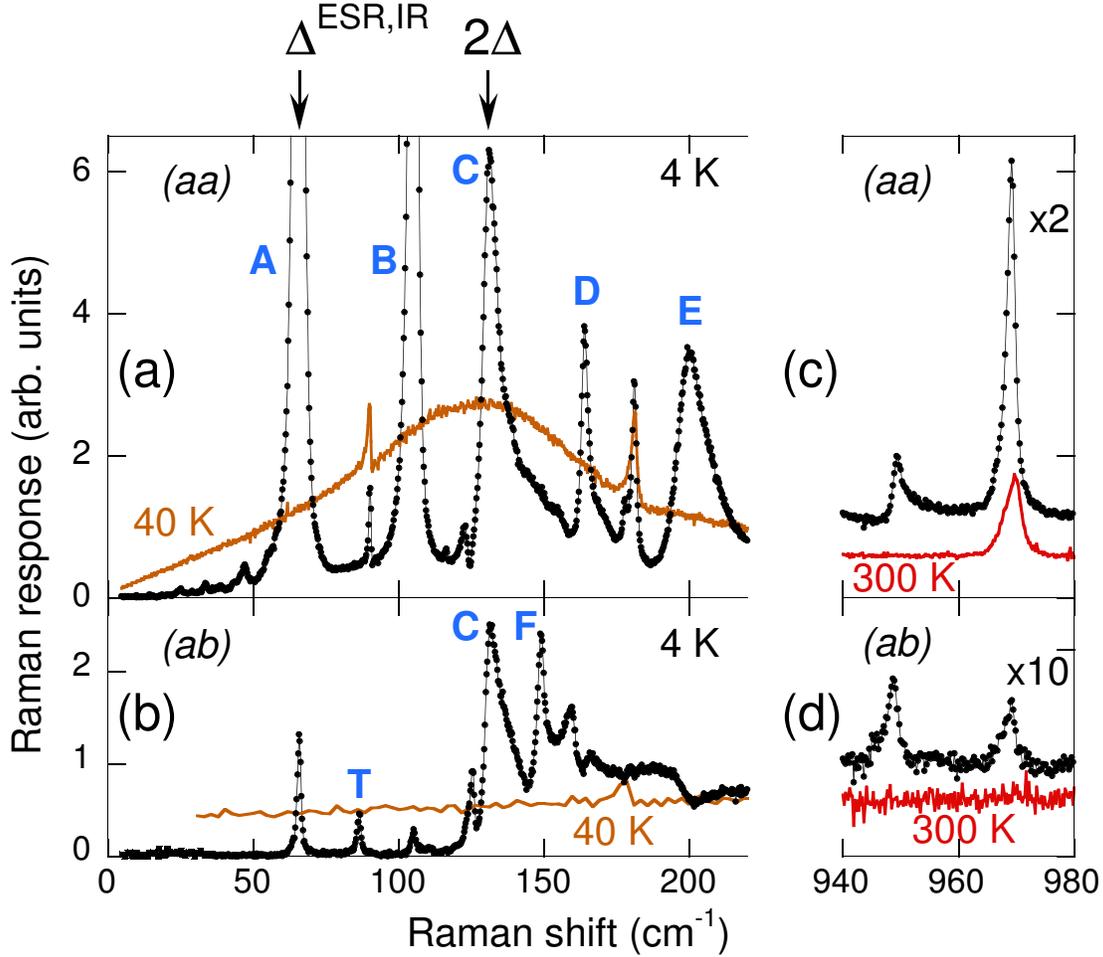,width=5.75in}
}
\caption{
Low energy Raman scattering spectra in $(aa)$ (upper panels) and $(ab)$ (lower panels) polarizations at 4, 40 and 300~K measured using the $\omega_{in} = 1.92$~eV excitation.
Below about 34~K many new resonances appear in the 0-200~\cm-1 region.
The arrows at the top correspond to the energies of the elementary triplet excitation determined by high resolution IR~\cite{RoomPRB04} and ESR~\cite{LutherJPSJ98} studies and twice that value.
The 968~\cm-1 phonon in parallel and cross polarization is shown on the right.
Panel (d) shows that below T$_{c}$ a folded counterpart appears which is seen in both $(aa)$ and $(ab)$ polarizations indicating a lower crystal symmetry.
}
\label{f518}
\end{figure}

Fig.~\ref{f517}d shows the ratio of the energy of lowest singlet bound state at the Brillouin zone center $m_{S = 0}$, with respect to the value of the elementary triplet gap $\Delta$.
The vertical line corresponds to the critical value of $\alpha$ at which a frustration induced gap opens up in the absence of any dimerization $\delta$.
At $\alpha_{c}$, the above ratio is $m_{S = 0} / \Delta = \sqrt{3}$ (the horizontal line refers to this value) and its value decreases with increasing frustration.
Note that at $\alpha = 0$, $m_{S = 0} / \Delta \approx 2$, and this is true irrespective of the chosen values for $\delta$ which means that it is the frustration which leads to finite binding energies of the singlet bound states.
One interesting point which is not shown in panel (d): there exists a singular point, the Majumdar-Ghosh point given by $(\delta,\alpha) \equiv (0,0.5)$ (see the discussion of Fig.~\ref{f59}), where $m_{S = 0} / \Delta$ equals unity.
At this point the single magnon and the two-magnon bound state are degenerate, which is a remarkable property of the Hamiltonian of Eq.~\ref{e53} \cite{TrebstThesis02}.

{\bf Experimental observations and discussion -- }
How is this discussion related to our experimental findings in \navo?
In Fig.~\ref{f518} we show several Raman spectra with the purpose of emphasizing the new excitations which appear below the phase transition.
We discuss first an aspect related to the crystal symmetry in the low temperature phase and illustrated in panels (c) and (d) of this figure.
We show there four Raman spectra, two in $(aa)$ polarization at 300 and 4~K, Fig.~\ref{f518}c, and two  in $(ab)$ polarization at the same two temperatures, Fig.~\ref{f518}d.
In the high temperature phase the V-O$_{rung}$ stretching mode \cite{GolubchikPopovicPopova97-99,KonstantinovicJPCM99} is present at 969~\cm-1 in parallel polarizations, but not in $(ab)$ geometry and, therefore it is a fully symmetric excitation of the $Pmmn$ space group. 
Below T$_{c}$ this mode acquires a second component, seen at 948~\cm-1, due to the in-plane unit cell doubling, i.e. we assign it to a folded phononic branch. 
More importantly, both the 968 and 949~\cm-1 excitations are seen not only in $(aa)$ but also in $(ab)$ polarization.
The simultaneous presence of these modes in both the diagonal and off-diagonal components of the Raman tensor indicates that in the low temperature phase the double reflection symmetry of the $Fmm2$ group is broken.
Our data confirm the resonant X-ray data \cite{GrenierPRB02JolyPRB03} which argues for a lower (monoclinic) symmetry below T$_{c}$.

We comment now on the spectra shown in Fig.~\ref{f518}a-b.
In panel (a) we show the emergence of new $(aa)$ polarized collective modes in the low temperature phase and several of them are indexed by letters: A (56.9\cm-1), B (105.0~\cm-1), C (131.2~\cm-1), D (164.0~\cm-1) and E (200.4~\cm-1).
In panel (b) we similarly marked the modes T (86.5\cm-1) and F (149.0\cm-1) seen in the $(ab)$ polarized spectra.
This list of modes does not exhaust all of the new excitations appearing in the low temperature phase but highlights a few of them which are of importance for the following discussion.
What is the origin of these new modes? Are they phonons, are they magnetic or are they some collective topological charge excitations of the type described in Ref.~\cite{AichhornPRB04}.
While not excluding the latter option, it seems that the observation of purely charge modes is less probable because it involves higher energies of the order intersite Coulomb interactions which are of the order of eV's \cite{MostovoySSC00andPRB02,AichhornPRB04}.
Moreover, the characteristic energy scales of these excitations (0-25~meV) can be better related either to magnetic excitations seen in ESR, far IR or INS experiments (see Fig.~\ref{f514}) or (in the context of existing lattice instabilities which drive the 34~K transition) to the folding of low energy zone boundary phonons in the high temperature phase. 
Accordingly, we will focus here on the possibility of observing phonons or single/multi-triplet magnetic excitations \cite{TrebstThesis02,GirshUnpublished}.

\begin{figure}[t]
\centerline{
\epsfig{figure=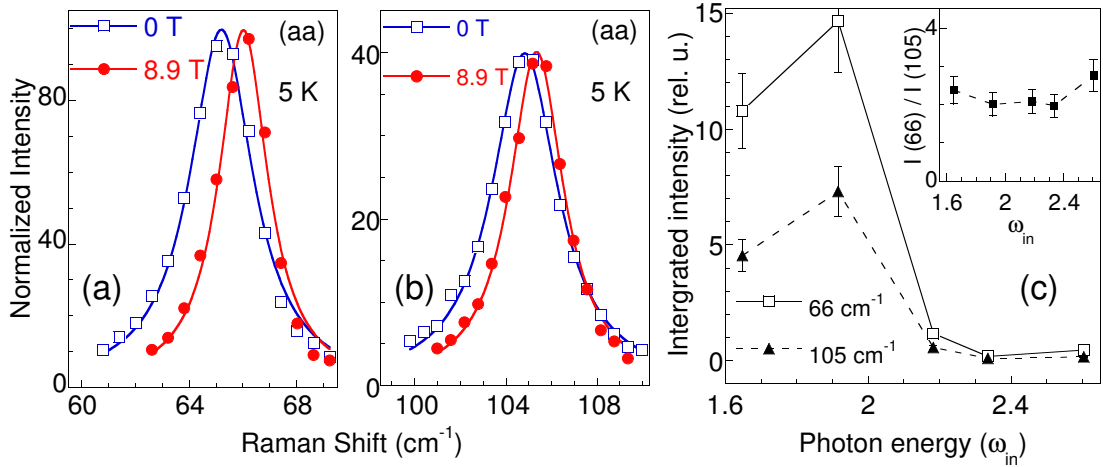,width=5.75in}
}
\caption{
Panels (a) and (b) display the T~=~5~K Raman spectra showing the magnetic field (in)dependence of the 66 and 105~\cm-1 resonances which appear below the transition.
In both cases a shift of about 0.9-1~\cm-1 was observed as the magnetic field was varied from 0 to 8.9~T.
(c) Integrated intensity of the 66 (empty squares) and 105~\cm-1 (filled triangles) modes at 5~K as a function of the incoming photon energies $\omega_{in}$.
The inset of this panel shows that the ratio of the integrated intensities of these two modes depend only slightly on $\omega_{in}$ (as opposed to the data shown in Fig.~\ref{f56}, the data in the main panel were not corrected for the optical properties of the material; however these effects cancel out in the ratio because these modes are very close in energy). 
}
\label{f519}
\end{figure}

We discuss here the spin excitations scenario.
This picture draws its appeal primarily from the fact that the Hamiltonian of Eq.~\ref{e53}, which seems to be a good description for the staggered spin configuration in the low temperature phase, allows for a multitude of magnetic bound states, both in the singlet and triplet channels, see Fig.~\ref{f517}.
First, we note that some of the new resonances observed below the transition are definitely magnetic.
In Fig.~\ref{f515} we showed that the 86.5~\cm-1 excitation, denoted by T in Fig.~\ref{f518}, is a $S~=~1$ magnon.
The singularity seen at 131.2~\cm-1 (16.27~meV) in Fig.~\ref{f518}a-b situated at twice the energy of the spin gap value of 65.5~\cm-1 (8.13~meV) and the continuum present from above this frequency as opposed to the clean gap seen for $\omega \leq 130$~\cm-1 (except for a few resonances) leaves little doubt that the mode C marks the onset of the two-triplet continuum of excitations.
How about the most prominent excitations denoted by A and B in Fig.~\ref{f518}a?
The similar resonance profile of these two excitations shown in Fig.~\ref{f519} indicates that they have a common origin.
It is interesting to remark that the energy of mode A coincides with the energy of the spin gap determined with high resolution in ESR and IR absorption data \cite{RoomPRB04,LutherJPSJ98}.
Can this excitation be the $k = 0$ elementary triplet?
Panels (a) and (b) in Fig.~\ref{f519} show that neither mode A nor mode B splits or shifts by more than 1~\cm-1 in applied magnetic fields up to 8.9~T (and we checked that this is true for any direction of the applied magnetic field).
Moreover, if these excitations were $S = 1$ triplets and the same spin-orbit interaction would insure the coupling to them as for mode T, one could not explain that the spectral weight of mode A is more than two orders of magnitude higher than that of mode T.

The arguments above seem to exclude the possibility that modes A and B are triplet excitations but the possibility that they are bound states in the singlet sector still needs to be explored.
The ratio $\omega_{B} / \Delta = 1.59$ appears to qualify mode B for a singlet bound state according to Fig.~\ref{f517}d.
How about mode A, can it be also a singlet bound state?
As discussed previously, a singlet bound state degenerate with the spin gap is only realized for a singular combination of parameters, $(\delta,\alpha) \equiv (0,0.5)$.
It deserves further argumentation why this set of values is realized in the low temperature phase of \navo.
However, assuming that this is the case, one could assign mode B to the folded counterpart of mode A.
In this scenario one could also argue that the magnetic frustration may be the driving force of the transition at 34~K in the following way \cite{GirshUnpublished}.
The tendency for zig-zag ordering increases the value of the NNN frustrating interaction $J_{2}$.
At 34~K, $\alpha  = J_{2} / J_{1}$ reaches the critical value $\alpha_{c}$ (see the caption of Fig.~\ref{f517}) and a gap opens in the magnetic excitation spectrum.
Further cooling will push $\alpha$ close to 0.5 and the singlet bound state gets very close (practically degenerate) to the spin gap.
Going into more details of this pictures some of the other excitations can be understood as combination modes or continuum edges.
For instance, based on energy considerations, one can say that F = A + B, H = B + B or that mode F in Fig.~\ref{f518}b is the A + T continuum edge.

However, the fact that the physics of the Majumdar-Ghosh point is realized for a singular combination of parameters makes this appealing scenario less probable.
The variety of selections rules which have been experimentally observed in Raman, ESR and IR data still needs to be explained.
The second possibility is to assign the main new features appearing below T$_{c}$ in the 0-200~\cm-1 energy range to folded phonons.
In fact, this scenario is supported by the small but finite shift observed in Fig.~\ref{f519}a-b.
This is what one expects in a simple two level system: one phonon, one close by magnon along with some off-diagonal matrix elements because of a non-vanishing spin-phonon interaction.
As expected, the shift of the 65.9~\cm-1 mode (of about 0.8~\cm-1) is slightly higher than the shift of the 105~\cm-1 mode (of abut 0.5~\cm-1) which is further apart in energy from the spin gap value. 
Based on lattice dynamical calculations and the relative temperature dependence of the spin gap seen in ESR data \cite{LutherJPSJ98} and that of the energies of measured IR active modes, a more recent study also argues for the phononic nature of modes A and B \cite{PopovaPRB02}.

We conclude this section by saying that many features of the transition at 34~K in \navo are still under debate and more than just a clarification is needed regarding the nature of the observed excitations or the effects of the electron-phonon interactions.

\section{Summary}

In this chapter we studied electronic excitations in \navo by Raman scattering and the transition taking place at T$_{c}$~=~34~K.
From the description of the crystal structure and the discussion of the electronic properties as seen by various techniques we showed that \navo can be thought of as an array of quasi one-dimensional (1D) two-leg ladders at quarter filling factor.
Two models were proposed to capture the essential physics of the magnetic degrees of freedom: one for the high temperature phase (Heisenberg chains with nearest and next nearest antiferromagnetic interactions, see Eq.~\ref{e51}), and another model for the low temperature phase (Heisenberg chains with dimerization and frustration, see Eq.~\ref{e53}).   

Three main topics were discussed.
The first was related to a broad continuum of excitations found in the 200-1500~\cm-1 range and peaked around 680~\cm-1 in parallel polarizations.
In literature a magnetic origin of this feature was ruled out and a scenario involving crystal field excitations was proposed instead.
The resonant Raman profile of this excitation, the polarization selection rules and the presence of its overtone in resonance conditions allowed us to conclude that the origin of this feature is magnetic.
More precisely, we proposed that it arises as a result of light coupling to multi-spinon Raman excitations.
Within this scenario we also argued for a scenario explaining the puzzling temperature dependence of the magnetic continuum in terms of an increasing role of next nearest neighbor frustration and in the context of a strongly fluctuating low temperature phase.

The second topic was related to the observation of a folded $S = 1$ magnetic mode which displayed very clear selection rules as a function of the magnetic field orientation, see Fig.~\ref{f515}.
We proposed that the coupling of the photon field to this excitation takes place $via$ the antisymmetric, Dzyaloshinskii-Moriya (DM), interaction which, in a simple dimer model, can also explain the observed selection rules: no splitting or shifts for magnetic fields parallel to the DM vector and the observation of two (upward and downward) dispersing branches for fields perpendicular to the DM vector.

Finally, we discussed the nature of several new resonances seen below T$_{c}$ and focussed on the possibilities that they are either folded phonons or singlet bound states of two triplet excitations.
In particular we emphasized the existence of two modes at 66 and 105~\cm-1, the first one being degenerate with one of the spin gap modes.
In the spin excitations scenario we proposed a model in which frustration has a determinant role.
However, in order to accommodate theoretical results with the observed energies, a very particular set of parameters characterizing the dimerized and frustrated spin chains (the Majumdar-Ghosh point) had to be invoked.
The above argument and the small energy shifts in magnetic fields up to 9~T suggested that the strong mode degenerate with the elementary triplet as well as the 105~\cm-1 mode are phonons.
The conclusion of our study is that many features of the spin/charge and lattice dynamics in \navo are still to be understood.


{\bf Acknowledgments --}
We acknowledge discussions and collaborations with B. S. Dennis, M. V. Klein, U.~Nagel,  T.~R\~{o}\~{o}m, A.M.~Sengupta and S.~Trebst. 
The crystals were provided by P.~Canfield.


\end{document}